%% file: main.tex
\documentclass[acmsmall,screen]{acmart}

\input{Text/definition}

\AtBeginDocument{%
  }

\begin{document}












\clearpage

\title{\tool: Automated Generation of Diverse Programming Problems for Benchmarking Code Generation Models}

\author{Simin Chen}
\email{simin.chen@UTDallas.edu}
\affiliation{%
  \institution{UT Dallas}
  \city{Dallas}
  \country{USA}
}
\author{Xiaoning Feng}
\email{fengxiaoning1746@link.tyut.edu.cn}
\affiliation{%
  \institution{Taiyuan University of Technology}
  \city{Taiyuan}
  \country{China}
}

\author{Xiaohong Han}
\email{hanxiaohong@tyut.edu.cn}
\affiliation{%
  \institution{Taiyuan University of Technology}
  \city{Taiyuan}
  \country{China}
}

\author{Cong Liu}
\email{congl@ucr.edu}
\affiliation{%
  \institution{UT Dallas}
  \city{Dallas}
  \country{USA}
}

\author{Wei Yang}
\email{wei.yang@utdallas.edu}
\affiliation{%
  \institution{UT Dallas}
  \city{Dallas}
  \country{USA}
}


\input{Text/abstract}






\maketitle

\input{Text/introduction}

\vspace{-3mm}
\input{Text/background}

\vspace{-3mm}
\input{Text/formulation}

\vspace{-3mm}
\input{Text/approach}

\vspace{-3mm}
\input{Text/setup}

\vspace{-3mm}
\input{Text/evaluation}

\input{Text/application}

\input{Text/discussion}

\input{Text/related}

\input{Text/conclusion}

\section{Replication Package}
Our code and data are available at \url{https://github.com/SeekingDream/PPM}

\clearpage
\bibliographystyle{plain}

\bibliography{acmart,xiaoning}

\end{document}

%% file: Text/definition.tex
\usepackage{booktabs}
\usepackage{amsfonts}
\usepackage{graphicx}
\usepackage{hyperref}
\usepackage{textcomp}
\usepackage{listings}
\usepackage{adjustbox}
\usepackage{xspace}    
\usepackage{multirow}
\usepackage{multicol}
\usepackage{xcolor} 
\usepackage{amsmath}
\usepackage{tcolorbox}
\usepackage[linesnumbered,boxed,ruled]{algorithm2e}
\usepackage{listings}
\usepackage{pifont}
\usepackage{enumitem}

\newcommand{\eg}{{\it e.g.,}\xspace}

\newcommand{\ie}{{\it i.e.,}\xspace}

\newcommand\figref[1]{Fig.~\ref{#1}}

\newcommand\tabref[1]{Table~\ref{#1}}

\newcommand\secref[1]{\S\ref{#1}}
\newcommand\equref[1]{Eq.(\ref{#1})}

\newcommand{\fakeparagraph}[1]{\noindent\textbf{#1.}}

\usepackage{wasysym}
\usepackage{algorithmic}

\definecolor{codegreen}{rgb}{0,0.6,0}
\definecolor{codegray}{rgb}{0.5,0.5,0.5}
\definecolor{codepurple}{rgb}{0.58,0,0.82}
\definecolor{backcolour}{rgb}{0.95,0.95,0.92}

\lstdefinestyle{mystyle}{
  backgroundcolor=\color{backcolour},   commentstyle=\color{codegreen},
  keywordstyle=\color{magenta},
  numberstyle=\tiny\color{codegray},
  stringstyle=\color{codepurple},
  basicstyle=\ttfamily\footnotesize,
  breakatwhitespace=false,         
  breaklines=true,                 
  captionpos=b,                    
  keepspaces=true,                 
  numbers=left,                    
  numbersep=5pt,                  
  showspaces=false,                
  showstringspaces=false,
  showtabs=false,                  
  tabsize=2
}

\def \lcgm{LCGM\xspace}
\def \lcgms{LCGMs\xspace}

\usepackage{nicematrix}
\def \tool{\texttt{PPM}\xspace}

%% file: Text/abstract.tex
\begin{abstract}
In recent times, a plethora of Large Code Generation Models (LCGMs) have been proposed, showcasing significant potential in assisting developers with complex programming tasks. Within the surge of LCGM proposals, a critical aspect of code generation research involves effectively benchmarking the programming capabilities of models.
Benchmarking LCGMs necessitates the creation of a set of diverse programming problems, and each problem comprises the prompt (including the task description), canonical solution, and test inputs. The existing methods for constructing such a problem set can be categorized into two main types: manual methods and perturbation-based methods. However, 
manual methods demand high effort and lack scalability, while also risking data integrity due to LCGMs' potentially contaminated data collection, and perturbation-based approaches mainly generate semantically homogeneous problems with the same canonical solutions and introduce typos that can be easily auto-corrected by IDE, making them ineffective and unrealistic.
Addressing the aforementioned limitations presents several challenges: (1) How to automatically generate semantically diverse Canonical Solutions to enable comprehensive benchmarking on the models, (2) how to ensure long-term data integrity to prevent data contamination, and (3) how to generate natural and realistic programming problems.
To tackle the first challenge, we draw key insights from viewing a program as a series of mappings from the input to the output domain. These mappings can be transformed, split, reordered, or merged to construct new programs. Based on this insight, we propose programming problem merging, where two existing programming problems are combined to create new ones.
In addressing the second challenge, we incorporate randomness to our programming problem-generation process. Our tool can probabilistically guarantee no data repetition across two random trials.
To tackle the third challenge, we propose the concept of a Lambda Programming Problem, comprising a concise one-sentence task description in natural language accompanied by a corresponding program implementation.  Our tool ensures the program prompt is grammatically correct. Additionally, the tool leverages return value type analysis to verify the correctness of newly created Canonical Solutions.
In our empirical evaluation, we utilize our tool on two widely-used datasets and compare it against nine baseline methods using eight code generation models. The results demonstrate the effectiveness of our tool in generating more challenging, diverse, and natural programming problems, comparing to the baselines.
\end{abstract}

%% file: Text/introduction.tex
\section{Introduction}

Recently, large code generation models (\lcgms) have garnered significant attention due to their remarkable performance in tackling complex programming problems. 
For instance, a prime example of such advancement is OpenAI's \textsf{CodeX} \cite{chen2021evaluating}, which can offer real-life help to software engineers and enhance their productivity.

Amidst the proposal of numerous \lcgms, a crucial aspect of code generation research revolves around effectively benchmarking the programming abilities of each model. 
One distinguishing characteristic sets benchmarking \lcgms apart when benchmarking their programming abilities, as opposed to benchmarking other machine learning models: 
To evaluate \lcgms, the model outputs (\ie the generated code snippets) must be executed with designated test inputs to observe its runtime behavior. 
In NLP, metrics like accuracy or BLEU score can quickly give an idea of the model's performance by comparing the output against a ground truth. However, in the context of code generation, such match-based metrics are insufficient for assessing the correctness of generated programs. 
Instead, determining correctness requires executing the generated program on designated test inputs and then conducting a functional equivalence comparison with a canonical solution.
Due to such distinct characteristics, a benchmark should encompass both the problem description and the canonical solution, along with test inputs for evaluating correctness.

To construct a programming problem dataset suitable for benchmarking purposes, one common approach involves employing manually-based methods \cite{humaneval}. However, we have identified two primary limitations associated with these methods.
\textcircled{1} Firstly, these methods necessitate a significant amount of human effort. This includes formulating precise programming task descriptions, creating accurate canonical solution implementations, and thoughtfully constructing comprehensive test inputs. 
These challenges result in a limited availability of evaluation data based on program execution. For instance, existing execution-based datasets like \textit{HumanEval}~\cite{chen2021evaluating} are notably scarce, featuring only a small number (a total of 164) of programming problems.
\textcircled{2} Secondly, these manually-based methods tend to generate concrete programming problem sets. Such concrete problem sets face the ongoing challenge of \textit{long-term data integrity}. In other words, once a programming problem dataset is published on the Internet for benchmarking purposes, future models may inadvertently utilize this dataset as part of their training data due to the extensive and sometimes indiscriminate collection of training data in \lcgms. 
This scenario may lead to unintended benchmark dataset leakage during the training of modern NLP models \cite{schaeffer2023pretraining}.

Another type of existing method belongs to perturbation-based methods \cite{recode}. These methods operate on the assumption that slight modifications to programming task descriptions should not alter the corresponding canonical solutions. Thus, they create programming problems by perturbing the programming task descriptions.
However, this method, while addressing certain automation limitations, encounters two significant drawbacks:
\textcircled{1}  Firstly, the programming problems generated by this method lack semantic diversity. 
Benchmarking a model using a semantic-homogeneous dataset may not accurately showcase the model's ability to effectively navigate the solution space. This is because it may fail to uncover edge cases or specific types of errors. Additionally, such datasets often exhibit varying levels of complexity, aiding in assessing the model's proficiency in handling tasks ranging from simpler to more intricate ones.
\textcircled{2} Secondly, such perturbation-based methods introduce unnatural and unrealistic alterations (e.g., typos, excessive blank lines) into the programming task descriptions. These alterations can be easily detected by modern Integrated Development Environments (IDEs), potentially diminishing the realistic of the generated problems.

We have identified three primary challenges in designing an automated method capable of effectively addressing the aforementioned limitations.
The first challenge is automating the generation of canonical solutions based on programming problem descriptions.
This challenge raises a dilemma regarding the potential obviation of the need for \lcgms if such a method exists that can autonomously create canonical solutions for any programming task.
The second challenge relates to ensuring long-term data integrity while also prioritizing transparency and public accessibility in benchmarking \lcgms.
Lastly, the third challenge involves generating programming problems that are natural and realistic.

\fakeparagraph{Our Idea} 
To tackle the challenges mentioned earlier, we have the following ideas:
\ding{182} To facilitate the automatic semantic programming problem generation, we make a crucial observation: A program's essence lies in its ability to map input from the domain to the corresponding output domain. Furthermore, we recognize the potential for using one program's output as another program's input. 
Building upon these insights, we introduce the concept of \textit{Programming Problem Merging} (\tool), wherein we combine two pre-existing programming problems to create a new one.
We employ this \textit{Programming Problem Merging} concept as the foundation of our method to generate new programming problems.
\ding{183} In the pursuit of ensuring long-term data integrity, our approach diverges from the creation of static programming problems. Instead, we propose injecting an element of randomness into our method. This entails defining an expansive random search space, allowing \tool to yield distinct programming problems with a high likelihood of avoiding repetition. While it's true that this element of randomness may introduce variability in benchmarking results, our contention is that repeated measurements can effectively mitigate this potential issue.
\ding{184} To address the challenge of generating natural and realistic programming problems, we introduce a novel concept known as the \textit{Lambda Programming Problem}. This \textit{Lambda Programming Problem} comprises a concise one-sentence task description in natural language, accompanied by a corresponding program implementation.
By incorporating the \textit{Lambda Programming Problem}, \tool can guarantee the grammatical correctness of newly generated task descriptions because both the seed task description and the task description in our \textit{Lambda Programming Problem} are grammatically correct. Furthermore, we conduct an analysis of the data type of a given seed programming problem and meticulously select an appropriate \textit{Lambda Programming Problem}. Consequently, \tool is equipped to ensure not only the novelty of canonical solutions but also their syntactical correctness.
Thus, the problems generated by \tool are more natural and realistic.

 \fakeparagraph{Implementation and Evaluation} 
We conducted comprehensive experiments to evaluate the effectiveness of \tool. Specifically, we applied \tool to two widely used real-world public datasets: \textit{HumanEval} and \textit{MBPP}. We benchmarked eight popular \lcgms that were trained on different corpora and featured diverse model architectures, sizes, and working mechanisms.
To gauge the performance of \tool, we compared it against nine state-of-the-art methods that are specifically designed for benchmarking code generation models. Our evaluation results demonstrated that \tool outperforms these methods, proving its high efficacy in generating test inputs that effectively degrade computation efficiency.

\begin{figure}[tbp]
    \centering
    \includegraphics[width=0.8\textwidth]{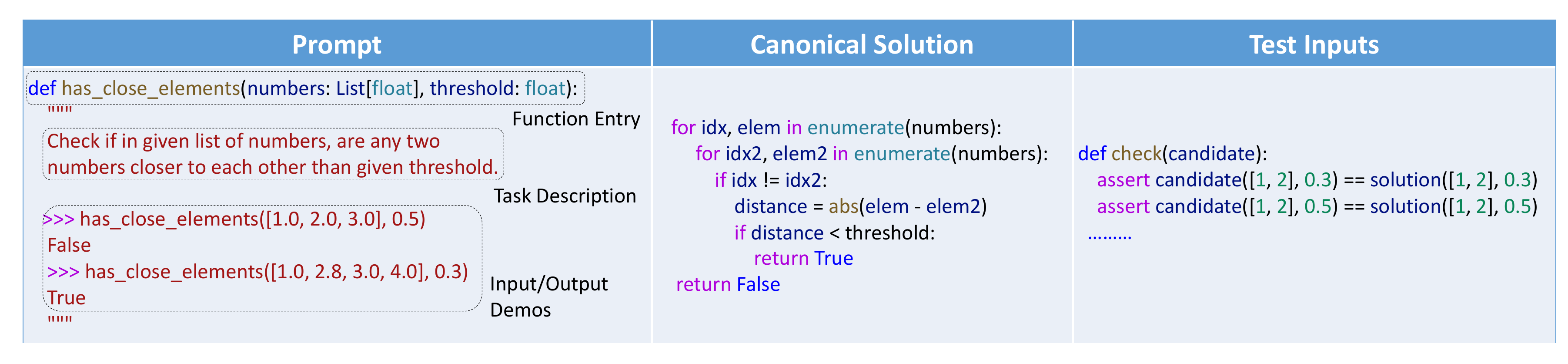}
    \vspace{-5mm}
    \caption{Programming problem example}
    \label{fig:demo}
\end{figure}

\fakeparagraph{Contributions} Our contributions are summarized as follows:

\begin{itemize}
    \item  We present the concept of \textit{Programming Problem Merging} (\tool), a novel methodology designed to create programming problems with diverse semantics. These problems are particularly valuable for benchmarking large code generation models.

    \item Based on the core principles of \tool, we offer two practical implementations that leverage lambda programming tasks: \textit{type-aware value transformation} (PPM-T) and \textit{pure value transformation} (PPM-V). These adaptations enable us to generate new semantic diverse programming problems.
    
    \item In our empirical evaluation of \tool, the results underscore its exceptional capacity to generate programming problems characterized by remarkable semantic diversity. This capability, in turn, sheds light on the inherent limitations of existing \lcgms, setting \tool apart from other methods that predominantly yield problems with homogeneous semantics. Furthermore, \tool excels in generating natural programming problems that maintain long-term data integrity, while providing stable benchmarking results.
\end{itemize}

%% file: Text/background.tex
\section{Background}
\label{sec:background}

\vspace{-1mm}
\subsection{Large Code Generation Models}
\label{sec:lcgm}


\fakeparagraph{Training of \lcgms} As the size of these models continues to grow, so does the training corpus they require. 
Due to the vastness of these training corpora for large code generation models, it becomes challenging to precisely identify the sources, for example, the training corpus of \textsf{CodeX} includes more than 54 million public software repositories \cite{humaneval}, and in some cases, the training data of the model remains unpublic \cite{gpt-4}.
Once the training corpus is crawled from the internet, the large code generation models are trained using the objective function, such as predicting the next tokens or predicting random masked tokens within the corpus. This training objective enables these models to learn from the vast data and improve their code generation capabilities.


\fakeparagraph{Evaluation of \lcgms} 
When comparing evaluation methods for machine learning models, assessing large code generation models reveals two unique characteristics.
(1). In evaluating ML models for NLP, common metrics like accuracy or BLEU score are used. However, these metrics fall short of capturing functionally equivalent programs, making it essential to assess the semantics of generated code by executing it on test inputs and comparing outputs with a reference solution.
(2). In contrast to traditional ML models that have predetermined training/testing datasets \cite{chen2023dark}, \lcgms don't have such splits due to their extensive training on a substantial portion of GitHub. The GitHub code corpus often already encompasses solutions to problems from various sources \cite{humaneval}. For example, the newly introduced \textit{APPS} dataset comprises over ten public repositories with solutions to Codeforces problems \cite{humaneval}. As a result, evaluating the programming capabilities of these models usually requires generating new custom programming problems.


\figref{fig:demo} showcases a programming problem example from the manually crafted \textit{HumanEval} dataset \cite{humaneval}, which is a widely used dataset for evaluating \lcgms. As shown in this figure, each programming problem comprises three parts: the prompt, the canonical solution, and a set of test inputs.
The prompt encompasses several crucial elements, namely a function definition that declares the entry point for the function, an extensive task description that provides detailed instructions on how to solve the problem, and illustrative input/output demonstrations.
The canonical solution is a manually crafted code snippet, carefully constructed to serve as the ground truth solution to the problem presented in the prompt.
The test inputs are designed to assess the correctness of program generated by code generation models when solving the problem defined in the prompt.

\vspace{-2mm}
\subsection{Programming Problem Generation Methods}


\fakeparagraph{Manually-based Methods} 
This category \cite{humaneval, mult-code, mbpp} encompasses methods that involve human expertise in crafting programming problems from scratch.
A noteworthy example is OpenAI's approach to assessing the code generation proficiency of Codex through the introduction of \textit{Human-Eval}, a comprehensive collection of 164 Python programming challenges, carefully designed and curated.

\fakeparagraph{Perturbation-based Methods} 
The previously mentioned manually-based methods demand significant human effort for the creation of dedicated programming problems and do not delve into the realm of robustness evaluation.
To overcome this limitation, a range of perturbation-based methods has emerged \cite{recode, li2022competition, icse-coderobustness}. These methods operate on the fundamental assumption that making slight modifications to the prompts in programming problems should not alter the corresponding canonical solutions.
One notable example of a perturbation-based method is \texttt{ReCode} \cite{recode}, which incorporates the concept of \textit{adversarial NLP}. It introduces a series of perturbations, such as token mutation, character mutation, and syntax mutation, to manipulate the prompts and assess the programming capabilities effectively.

\input{Table/existing}

\fakeparagraph{Optimal Design Space \& Limitation of Existing Methods}
We have outlined four key features of an ideal method for creating programming problems, as summarized in \tabref{tab:limitation}.
\ding{182} \textit{Automation}: The method must be automated to eliminate the necessity for extensive manual dataset creation. Manual methods are labor-intensive and cannot produce scalable datasets.
\ding{183} \textit{Semantic Diversity}: The method should generate programming problems that exhibit semantic diversity. Failing to do so, by generating a set of semantically uniform problems, would result in an incomplete evaluation of the programming capabilities of \lcgms, potentially yielding falsely elevated results.
\ding{184} \textit{Long-Term Integrity}: The method should include mechanisms to resist data leakage, especially for maintaining long-term data integrity. Given \lcgm's reliance on internet-sourced training data, pre-existing manually crafted datasets could unintentionally become part of future \lcgm models, rendering them inaccessible for benchmarking purposes.
\ding{185} \textit{Naturalness}: The method should generate natural and realistic programming problems. Modern Integrated Development Environments (IDEs) excel at detecting typos and grammatical errors, rendering grammar-incorrect programming problems unrealistic and impractical. 
Regrettably, as indicated in \tabref{tab:limitation}, existing methods do not comprehensively support these four essential features. Hence, there is a compelling need to design a new method that addresses these limitations.


%% file: Table/existing.tex
\begin{table*}[tbp!]
  \centering
  \caption{Feature Support in Existing Work within the Optimal Design Space. \CIRCLE represents supported, \LEFTcircle represents partial supported, \Circle represents not supported, and  $-$ represents not applicable.}
  \vspace{-3mm}
  \resizebox{0.88\textwidth}{!}{
    \begin{NiceTabular}{lccccl}
    \CodeBefore
    \rowcolors{1}{}{blue!8}
        \Body
    \toprule
    \toprule
    \textbf{Methods} & \textbf{Automation} & \textbf{Semantic Diversity} & \textbf{Long-Term Integrity} & \textbf{Naturalness} & \textbf{Representative Works} \\
    \midrule
    \textbf{Hand-written} & \Circle  & $-$ & $-$ & \CIRCLE & Human-Eval, MBPP \\
    \textbf{Description Perturbation} & \CIRCLE & \Circle  & \LEFTcircle  & \LEFTcircle  &  \texttt{Character Mutation}, \texttt{Token Mutation}\\
    \textbf{Sytanx Perturbation} & \CIRCLE & \Circle  & \LEFTcircle  & \CIRCLE &  \texttt{Insert Blank Line} \\
    \textbf{Demo Perturbation} & \CIRCLE & \Circle  & \LEFTcircle  & \CIRCLE & \texttt{Add Demo}, \texttt{Replace Demo}  \\ \hline 
    \textbf{Ours} & \CIRCLE & \CIRCLE & \CIRCLE & \CIRCLE &  \tool \\
    \bottomrule
    \bottomrule
    \end{NiceTabular}%
  }
  \label{tab:limitation}%
\end{table*}%

%% file: Text/formulation.tex
\section{Challenges \& High-level Solutions }
\label{sec:challenge}

\fakeparagraph{Challenge 1: Automated Generation of Semantic Diverse Problems}  
Generating programming problems with diverse semantics automatically poses a significant challenge. This challenge arises from the need for unique canonical solutions. 
Automatically generating correct canonical solutions for any programming problem presents a considerable challenge. 
The reason is that if we already have an automated method capable of creating canonical solutions, the existing \lcgm models would be redundant. Such an automated method could then be used to directly assist developers in generating programs, eliminating the need for developing different \lcgms. 

\noindent\textbf{\textit{Solution 1}}: To tackle this challenge and create programming problems that exhibit semantic diversity, we introduce a novel idea, \textit{Programming Problem Merging}. wherein we combine two pre-existing programming problems to create a new one. The newly created programming problem will possess distinct semantics compared to the original two problems. Additionally, we can leverage the canonical solutions from the two problems to automatically derive the correct canonical solution for the merged problem.


\fakeparagraph{Challenge 2: Dilemma of Long-Term Integrity and Public Available}
Any specific, concrete programming problems have inherent limitations of long-term integrity once they are published on the Internet due to the specific training data collection mechanism in \lcgms (\secref{sec:lcgm}).
However, benchmarking \lcgms necessitates the benchmarks being transparent and publicly accessible—a requirement that seemingly contradicts the concept of long-term integrity.

\noindent\textbf{\textit{Solution 2}}: 
In response to this challenge, instead of generating specific, concrete programming problems, we introduce a methodology that is attuned to randomness, allowing us to generate diverse programming problems. The search space in our random parameter search is substantial, minimizing the likelihood of producing identical programming problems. By adopting this randomness-aware idea, our method ensures Long-term integrity. In the event that a particular set of concrete programming problems becomes exposed, we can confidently generate non-repeating, randomized problems for evaluation, significantly reducing the risk of data compromise.

\fakeparagraph{Challenge 3: Natural and Realistic Problems}
The final challenge pertains to generating natural and realistic programming problems, which require grammatical correctness in the programming task description. However, directly combining two existing complex programming task descriptions often results in grammar errors and makes the new problem unnatural.

\noindent\textbf{\textit{Solution 3}}:
To overcome this challenge, we introduce the concept of a \textit{lambda programming task}. 
This involves a concise one-sentence description aimed at manipulating the output of an existing programming task description. Since both the original programming task description and our proposed \textit{lambda programming task} are grammatically correct, we can guarantee the correctness of the newly generated programming problem.

%% file: Text/approach.tex
\section{Approach}


\input{Approach/formulation}

\input{Approach/overview}

\input{Approach/detail}

\input{Approach/test}

%% file: Approach/formulation.tex
\vspace{-1mm}
\subsection{Problem Formulation of Programming Problem Merging}
\label{sec:fomulation}


In formal terms, let us examine a seed programming problem, represented as $\mathcal{P} = <P, S, T>$, where: (1) input prompt $P$ symbolically expressed as $P = <f, t, d>$. In this context, $f$ denotes the function entry declaration, $t$ represents the task description and $d$ encompasses input/output demonstrations; (2) $S$ represents the canonical solution of the problem; (3) Test inputs $T$ validate whether \lcgm generated program is correct by comparing the generated solution with the canonical solution.

Our objective is to search for a concise one-sentence natural language description $\phi$ along with its corresponding implementation, $\lambda$. 
The purpose of $\phi$ is to provide clear instructions on how to handle the output value of the seed canonical solution, and $\lambda$ translates these instructions into code implementation, aligning with the one-sentence natural language description $\phi$. 
Consequently, the newly generated programming problem can be represented as $\mathcal{P}_{new} = < \phi \circ P, \lambda \circ S(\cdot), T>$.

%% file: Approach/overview.tex
\vspace{-2mm}
\subsection{Design Overview}

\label{sec:overview}

\begin{figure}
    \centering
    \includegraphics[width=0.9\textwidth]{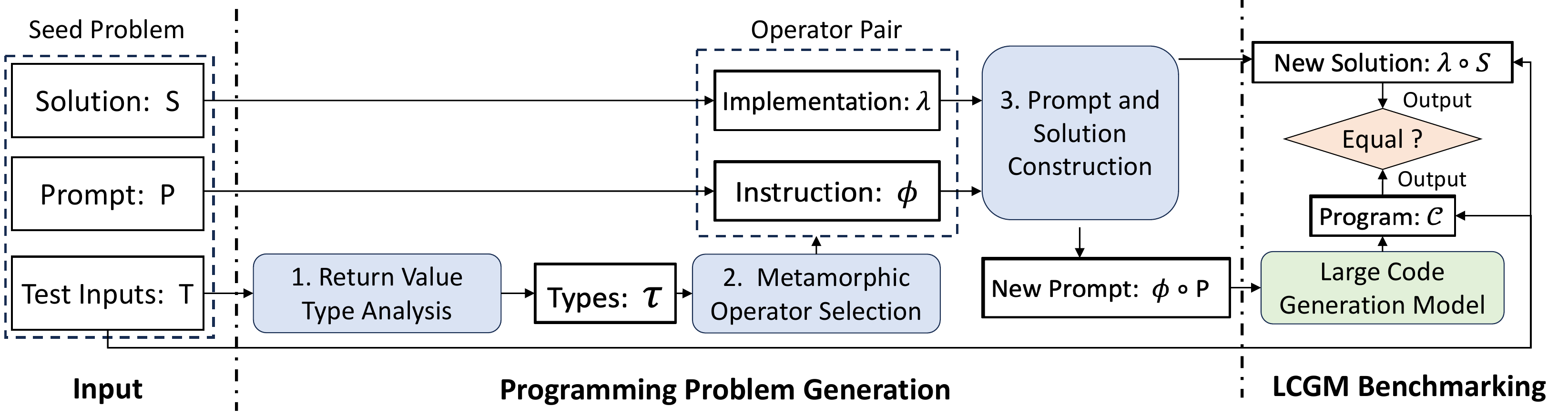}
    \vspace{-5mm}
    \caption{Design overview of \tool}
    \vspace{-5mm}
    \label{fig:overview}
\end{figure}

The design overview of \tool is shown in \figref{fig:overview}, \tool accepts a seed problem as input and generates a new and semantically-different programming problem by performing three main steps: 
\begin{enumerate}
    \item \textit{Return Value Type Analysis}. In the first step, \tool performs return value analysis for each seed programming problem. It collects the returned values of the canonical solution and extracts their corresponding data types.
    
    \item \textit{Metamorphic Operator Selection}. Following the analysis of return values, \tool proceeds to select an appropriate metamorphic operator based on the extracted data type. These operators, drawn from a predefined set (\tool provides templates for user to expand this set), consist of paired transformations for both the prompt ($\phi$) and the canonical solution ($\lambda$). Additionally, \tool randomly generates the parameters required for these selected transformations.
    
    \item \textit{Prompt and Solution Construction}. Building upon the chosen metamorphic operator, \tool constructs a new programming problem by skillfully merging the prompt description with the solution, creating a new programming problem.
    
\end{enumerate}

%% file: Approach/detail.tex


\vspace{-2mm}
\subsection{Return Value Type Analysis}

The fundamental concept behind \tool is the fusion of two programming problems to generate a novel programming challenge with semantic diversity. Specifically, our aim is to leverage the output of a seed programming problem as input for our \textit{lambda programming problem}.
To select a \textit{lambda programming problem} that avoids the introduction of type errors, the initial step necessitates the compilation of return value types from the seed problem. It's worth highlighting that specific strongly typed programming languages (such as C/C++ and Java) explicitly specify return types within function entry declarations.
Regrettably, the prevailing trend in existing programming problem benchmarks leans heavily toward Python. Python, being one of the most widely used programming languages in the field of machine learning, is inherently weakly typed, lacking explicit return value type definitions.

In order to gather return value types in weakly typed programming languages, our approach entails the acquisition of actual return values through execution, followed by an analysis to ascertain their respective return types.
Delving into further detail, our process begins with the definition of basic data types. 
Following this, we execute the canonical solution on the test inputs associated with the seed problem and subsequently analyze for potential complex data types.

\fakeparagraph{Basic Data Type Definition} 
We have defined four fundamental built-in data types in Python, which serve as our core data types: \textit{int}, \textit{float}, \textit{string}, \textit{boolean}. These defined basic data types encompass a wide range of programming language constructs and can be extended for use in other programming languages.

\fakeparagraph{Data Type Abstraction}
Given the seed programming problem, we begin by subjecting all test cases from the test inputs to the canonical solution to execute and gather the output values. Subsequently, we analyze these output values and extract basic data types from the collected set. 
In addition to basic data types, a programming problem may produce other enumerable data types (\eg lists or tuples in Python). Therefore, we propose a recursive algorithm to abstract the data types of the return values.
Our data type abstraction algorithm, as depicted in Algorithm \ref{alg:alg}, takes a list of returned values as input and iterates through each value in the list. If the value's type belongs to our predefined set of basic data types, we include that data type in our collection (Line \ref{alg:basic}). If the data value belongs to enumerable data types, we convert the value into a list and determine the types of all its elements (Line \ref{alg:complex}).
Through this process, we are able to gather all basic data types present within the returned values.

\input{Alg/alg1}


\subsection{Metamorphic Transformation Operator Selection}

\input{Table/op}

Once we have gathered the set of return value types, denoted as $\Vec{\tau}$, from the canonical solution, we proceed by selecting a data type at random from this set. Additionally, we randomly choose a transformation operator from our predefined set of operators.
Although \tool can support any user-defined operators, as a proof of concept, we introduce two types of transformation operators: \textit{data type-aware value transformation} and \textit{pure value transformation}. Our predefined operator set is detailed in \tabref{tab:op}, where column \textit{src\_type} represents the randomly selected source data type, \textit{tgt\_type} signifies the target data type to which we aim to transform, \textit{Natural Language Description} provides a concise one-sentence natural language description denoted as $\phi$, which we append to the original task description, and \textit{Implementation} column presents the corresponding code implementation $\lambda$ based on the aforementioned natural language description.

In the \textit{Implementation} column, it's worth noting that the lambda expression requires two parameters: $x$, which represents the output value of the seed canonical solution, and $offset$, a randomly generated variable. The introduction of this random variable $offset$ in the metamorphic transformation operators serves the purpose of injecting randomness into the newly generated programming problem. This injection of randomness is crucial for achieving long-term problem integrity.
Furthermore, the space for the random variable $offset$ is user-configurable. By allowing users to define a sufficiently large search space for $offset$, \tool can effectively generate unique programming problems with a high probability of distinctiveness.

We present a theoretical analysis elucidating the probability of each proposed operator generating identical programming problems across two distinct trials on our website.

\vspace{-2mm}
\subsection{Prompt and Canonical Solution Construction}

\label{sec:construction}

After generating a randomized metamorphic operator, comprising a one-sentence natural language description $\phi$ and its corresponding implementation $\lambda$ in the previous step, \tool proceeds with the following steps to craft a novel programming problem.

\fakeparagraph{Prompt Construction} 
In the prompt creation phase, \tool initially concatenates the one-sentence natural language description with the original task description, resulting in a new task description. 
Subsequently, \tool collects output demonstrations from the input/output demonstrations associated with the prompt and inputs them into the corresponding implementation to generate new output demonstrations. 
Through the combination of the original function entry declaration, the newly crafted task description, and the new generated output demonstrations, \tool creates a new prompt.

\fakeparagraph{Canonical Solution Construction}
To generate a new canonical solution, \tool concatenates the implementation $\lambda$ with the original solution construction $S$. 
In essence, we treat the output of the original solution construction as the input for the lambda implementation.
It's important to note that our one-sentence natural language description $\phi$ specifies only one data type. 
Consequently, we apply a filter to ensure that the data type aligns with our natural language description before feeding it into our lambda implementation.
\tool regards the ordered sequence of function calls within $S$ and $\lambda$ as our newly defined canonical solution.



%% file: Alg/alg1.tex
    \begin{algorithm}[tbp!]
\caption{Data Type Abstraction Algorithm.  $\quad$ \texttt{Abstract} ($\cdot$)} 
\label{alg:alg}

\begin{flushleft}
 {\bf Input:} Value list $\mathcal{V}$. \\
 {\bf Output:} Set of possible data types of returned values $\vec{\tau}$. \\
\end{flushleft}

\begin{algorithmic}[1]


\STATE $\vec{\tau}$ = \{$\;$\}  $\quad\quad\quad\quad\quad\quad\quad\quad\quad\quad\quad\quad\quad\quad$ // Initialize possible data types as an empty set.

\FOR {each v $\text{in} \; \mathcal{V}$}
    \IF {\texttt{Type}(v) $\in$ Basic Types}
        \STATE \label{alg:basic} $\vec{\tau} = \vec{\tau} + \texttt{Type}(v)$  $\quad\quad\quad\quad\quad\quad\quad\quad$ // Add current basic data type to the set.
    \ELSIF{\texttt{Type}(v) $\in$ Enumerable Types}
        \STATE \label{alg:complex} $\vec{\tau}  = \vec{\tau} + \texttt{Abstract}(\texttt{ToList}(v))$   $\quad\quad$ 
        // Iteratively add each element type into the set.
    \ENDIF
\ENDFOR
\STATE return $\vec{\tau}$
\end{algorithmic}
\end{algorithm}

%% file: Table/op.tex
\begin{table*}[t!]
    \centering
    \caption{Metamorphic Transformation Operator Set in \tool.}
    \vspace{-3mm}
    \label{tab:op}
    \resizebox{\linewidth}{!}{%
    \begin{NiceTabular}{p{3.5em}p{3.5em}ll}
        \CodeBefore
            \rowcolors{2}{}{blue!8}
        \Body
    \toprule
    \toprule
    \textbf{src\_type} & \textbf{tgt\_type} & \multicolumn{1}{l}{\textbf{Natural Language Description}} & \textbf{Implementation} \\
    \midrule
    \multirow{3}[2]{*}{\textbf{Int}} & \textbf{float} & Change all {src\_type} type values of the return values to {tgt\_type} type, and add {offset}. & $\lambda(x)$: float(x) + offset \\
          & \textbf{string} & Change all {src\_type} type values of the return values to {tgt\_type} type, return the string value of answer + {offset}. & $\lambda(x)$: str(x + offset) \\
          & \textbf{boolean} & Change all {src\_type} type values of the return values to {tgt\_type} type, change all odd results to {offset}, and all even results to {not offset}. & $\lambda(x)$: offset if x \% 2 else not offset \\
    \midrule
    \multirow{3}[2]{*}{\textit{\textbf{float}}} & \textit{\textbf{int}} & \textit{Change all {src\_type} type values of the return values to {tgt\_type} type, keep the integer part of the result plus {offset}.} & $\lambda(x)$: str(x + offset) \\
          & \textit{\textbf{string}} & \textit{Change all {src\_type} type values of the return values to {tgt\_type} type, return the string value of answer + {offset}.} & $\lambda(x)$: int(x) + offset \\
          & \textit{\textbf{boolean}} & \textit{Change all {src\_type} type values of the return values to {tgt\_type} type, if the answer is larger than 0.0, return {offset}, else return {not offset}.} & $\lambda(x)$: offset if x > 0.0 else not offset \\
    \midrule
    \multirow{3}[2]{*}{\textbf{string}} & \textbf{int} & Change all {src\_type} type values of the return values to {tgt\_type} type, and return the length of the string plus {offset}. & $\lambda(x)$: len(x) + offset \\
          & \textbf{float} & Change all {src\_type} type values of the return values to {tgt\_type} type, and return the length of the string plus {offset}. & $\lambda(x)$: len(x) + offset \\
          & \textbf{boolean} & Change all {src\_type} type values of the return values to {tgt\_type} type, change all odd-length strings to {offset}, and all even-length strings to {not offset}. & $\lambda(x)$: offset if len(x) \% 2 else not offset \\
    \midrule
    \multirow{3}[2]{*}{\textbf{boolean}} & \textbf{int} & Change all {src\_type} type values of the return values to {tgt\_type} type,  and add {offset} & $\lambda(x)$: int(x) + offset \\
          & \textbf{float} & Change all {src\_type} type values of the return values to {tgt\_type} type,  and add {offset} & $\lambda(x)$: int(x) + offset \\
          & \textbf{string} & Change all {src\_type} type values of the return values to {tgt\_type} type, and change True to {offset}, and False to {chr(ord(offset) + 1)}. & $\lambda(x)$: offset if x else chr(ord(offset) + 1) \\
    \midrule
    \midrule
    \textbf{int} & \textbf{int} & For all {src\_type} type values in the return results, increase each value by {offset}. & $\lambda(x)$: x + offset \\
    \textbf{float} & \textbf{float} & \multicolumn{1}{l}{For all {src\_type} type values in the return results, increase each value by {offset}.} & $\lambda(x)$: x+ offset \\
    \textbf{string} & \textbf{string} &   For all {src\_type} values in the return results, map each character in the {src\_type} value to the character whose ASCII number is the current ASCII value plus {offset}. & $\lambda(x)$: ''.join([chr(ord(char) + offset) for char in x]) \\
    \textbf{boolean} & \textbf{boolean} & For all {src\_type} values in the return results, invert True to False and False to True. & $\lambda(x)$: not x \\
    \bottomrule
    \bottomrule
    \end{NiceTabular}%
    }
\end{table*}

%% file: Approach/test.tex
\vspace{-2mm}
\subsection{Benchmarking Stage}

\label{sec:benchmark}

To employ our newly created programming problem for benchmarking \lcgms, we first feed the new prompt $\phi \circ P$ to the \lcgm and collect the generated program $C(\cdot)$.
Then we evaluate whether the equation holds for all test inputs $\mathcal{C}(x) == \lambda \circ S(x) \quad \forall x \in T$. If the equation holds, then the generated program $C(\cdot)$ is the correct one.
By measuring the accuracy of the generated programming problem set, \tool can benchmark \lcgms.

%% file: Text/setup.tex
\section{Experimental Setup}

\input{Setup/rq}

\input{Setup/dataset}

\input{Setup/model}

\input{Setup/baseline}

\input{Setup/metric}

\input{Setup/process}

\input{Setup/implementation}

%% file: Setup/rq.tex
We present an empirical evaluation and aim to address the following research questions:
\begin{enumerate}[label=\textbf{RQ\arabic*}]
    \item \label{rq:diversity} \textbf{Diversity:} 
      How diverse are the programing problems generated by \tool? 

    \item \label{rq:effectiveness}\textbf{Effectiveness:}
    How effectively the generated programming problems reveal the issues and limitations of existing \lcgms?

    \item \label{rq:naturalness} 
     \textbf{Naturalness:}
    How  natural and realistic are the programming problems generated by \tool ? 
    
    \item \label{rq:sensitivity} 
     \textbf{Stability:}
    Can \tool consistently produce stable benchmarking results despite randomness and under varying hyperparameters?

\end{enumerate}

%% file: Setup/dataset.tex
\vspace{-2mm}
\subsection{Datasets}

We perform experiments using two datasets: \textit{HumanEval} \cite{humaneval} and \textit{MBPP-Sanitized} \cite{mbpp}.

\fakeparagraph{(1) \textit{HumanEval}} The \textit{HumanEval} dataset is an open-sourced benchmark proposed by OpenAI for evaluating the code-generation ability of pre-trained \lcgm. It comprises 164 Python programming problems, each of which includes a prompt, a canonical solution, and a set of test inputs. The prompt consists of a natural language description of the problem to be solved, a function definition, and several input/output pairs.

\fakeparagraph{(2) \textit{MBPP-Sanitized}} The dataset utilized for our experiment comprises 427 Python programming questions collected from crowdsourcing. This dataset is considered a zero-shot dataset since it lacks any input/output demonstrations in its prompts.
To enhance the experiment's effectiveness, we applied a prompt format modification. In detail, each problem was processed by adding a function header and converting the natural language instructions into function docstrings. 



%% file: Setup/model.tex
\vspace{-2mm}
\subsection{Pre-trained Code Generation Models}

In this work, we perform a comprehensive evaluation of  \tool on eight popular public \lcgm. The selected \lcgm are diverse in terms of model architecture, model size, and training methods. 
\textbf{(1) \textsf{CodeGen} Family} This type of model \cite{nijkamp2022codegen} belongs to a family of autoregressive language models that undergo pretraining on code data. Our experimentation encompasses the utilization of two variants within the open-source \textsf{CodeGen} model family, scaling 6B-half and 2B separately.
\textbf{(2) \textsf{CodeGen2} Family} \textsf{CodeGen2}~\cite{nijkamp2023codegen2} is an upgraded version of the \textsf{CodeGen} family, which falls under the category of autoregressive language models. It builds upon the capabilities of \textsf{CodeGen} and introduces additional features such as code filling. 
\textbf{(3) \textsf{InCoder} Family} The \textsf{Incoder} model~\cite{fried2022incoder} is available in two sizes: 1B and 6B. The decoder-only transformer utilizes a causal-masked objective during training to effectively handle code generation.
\textbf{(4) \textsf{SantaCoder} Family} \textsf{SantaCoder}~\cite{allal2023santacoder} is a model that utilizes a multi-head attention mechanism as its primary model architecture. Despite having a relatively compact size with 1.1 billion model parameters, \textsf{SantaCoder} delivers exceptional performance surpassing that of other models with similar parameter sizes. 
\textbf{(5) \textsf{PloyCoder}} This model utilizes the \textsf{GPT2} architecture and underwent training on an extensive dataset of 249 GB of code encompassing 12 programming languages. It comprises an impressive 2.7 billion parameters trained over 100,000 to 150,000 steps.

%% file: Setup/baseline.tex
\vspace{-2mm}
\subsection{Comparison Baselines}

We compare \tool against night baseline methods designed for creating programming problems.
\textbf{(1) \texttt{Base}} This approach involves directly utilizing the prompts provided in the manually crafted dataset to generate code snippets without any modification. 
\textbf{(2) \texttt{Add Demo}} This method involves the random selection of a test case from the test inputs. It proceeds by calculating the expected output through the application of the provided canonical solution. Subsequently, it adds this new input/output case into the seed prompt, alongside the original input/output pairs, to create a new prompt.
\textbf{(3) \texttt{Remove Demo}} Similar to the prompt crafting method in \texttt{Add Demo}, this method randomly removes an input/output demo from the original prompt.
\textbf{(4) \texttt{Replace Demo}} Similar to the prompt crafting method in \texttt{Add Demo}, this method randomly replaces an input/output demo in the original prompt with the one in the test inputs.
\textbf{(5) \texttt{Token Mutation}} This method is widely used in existing work to evaluate the robustness of natural language processing (NLP) models \cite{recode}. Specifically, this method randomly replaces a token in the task description with another token using token substitution. 
\textbf{(6)\texttt{Character Mutation}} This method entails the random alteration of a token within the natural language task description at the character level. In our experiment, we examine four distinct types of character mutation operators: (1) neighbor character swap, (2) random character insertion, (3) random character deletion, and (4) homoglyph character replacement. 
\textbf{(7) \texttt{FuncName Mutation}} This method introduces perturbations to function names within the function entry declaration using the following operators: (1) CamelCase transformation, which involves altering function names between camel-case and snake-case (\eg ``findCharLong'' to ``find\_char\_long''). (2) Application of all character mutation operators found in the \texttt{Character Mutation} method.
\textbf{(8) \texttt{Empty Line Insertion}} This method~\cite{recode}  involves the insertion of an empty line into the seed prompt to generate a new prompt variant. 
\textbf{(9) \texttt{CommSyntax}} This method~\cite{recode} transforms the syntax of the docstring section, including the task description and input/output demonstrations, from its original format (\eg """docstring""") into comment syntax (\eg $\#$docstring).


%% file: Setup/metric.tex
\vspace{-2mm}
\subsection{Evaluation Metrics}

\input{Setup/Metric/diversity}

\input{Setup/Metric/effectiveness}

\input{Setup/Metric/natural}

\input{Setup/Metric/stable}

%% file: Setup/Metric/diversity.tex
\fakeparagraph{Diversity Metrics}  To assess the diversity of the generated programming problems,  we conduct a measurement encompassing both the prompt and canonical solution aspects. In evaluating prompt diversity, we employ two metrics: (1) \textit{BLEU score} and (2) \textit{Semantic Similarity}. As for assessing diversity in canonical solutions, we rely on the \textit{Different Implementation Rate} metric.
\textit{BLEU score} is widely employed as a quantitative measure of the linguistic diversity of natural language sentences. 
The \textit{BLEU score} is formally defined in \equref{eq:bleu}, 
\begin{equation}
\label{eq:bleu}
 \text{BLEU} = \text{BP} \times \exp\left(\sum_{n=1}^N w_n \log p_n\right)  
        \text{BP} =
\begin{cases}
1 & \text{if c} \geq \text{r} \\
\exp^{(1 - \frac{\text{r}}{\text{c}})} & \text{otherwise} 
\end{cases}
\end{equation}
where BP is the brevity penalty, which is a correction factor that accounts for the difference in length between the candidate and reference sentences. 
Following previous works, we use \textit{BLEU-4}, where $N$ is set to 4, and $w_n$ is set to 0.25 for all $n$ (uniform weighting).
Lower \textit{BLUE scores} indicate a greater mismatch to the original problems and reflect a higher level of diversity.
\textit{Semantic Similarity (SemSim)} serves as a crucial metric in evaluating the semantic diversity of generated programming problem prompts. It quantifies the similarity in semantic meaning between the generated problems and seed problems. The \textit{Semantic Similarity} is computed using the following equations:
\begin{equation}
\label{eq:sim}
\text{Semantic Similarity} = \frac{1}{N} \sum_{i = 1}^{N} \text{cos}(\text{Embed}(P_i), \; \text{Embed}(P_i'))
\end{equation}
In this equation, $\text{Embed}(\cdot)$ represents the embedding function that projects sentences to numeric vectors. $P_i$ and $P_i'$ are the $i^{th}$ original problem and generated problem descriptions, respectively. The function $\text{cos}(\cdot)$ measures the cosine similarity between two numeric vectors.
A lower \textit{Semantic Similarity} score suggests greater diversity among the generated programming problems, as they exhibit distinct variations in semantic meaning compared to the original problems. 
\textit{Different Implementation Rate (DiffImp).} In order to assess the diversity of the canonical solutions, we employ the \textit{Different Implementation Rate} metric. This metric is calculated by determining the percentage of generated programming problems that employ distinct canonical solutions compared to the seed programming problems.

%% file: Setup/Metric/effectiveness.tex
\fakeparagraph{Effectiveness Metrics}  
To evaluate the programming capability of each \lcgm with the crafted programming problems, we adopt a well-established methodology used in previous research \cite{humaneval}. 
For this assessment, we utilize the \textit{Pass@k} metric, which serves as a measure of the functional correctness of the generated code by executing test cases. The \textit{Pass@k} metric is defined as follows:
\begin{equation}
\label{eq:pass}
\text{Pass@k} = \mathbb{E}_{\text{problems}} \left[1 - \frac{\binom{n - c}{k}}{\binom{n}{k}}\right]
\end{equation}
In the equation, the symbol $\mathbb{E}_{\text{problems}}$ represents the expected value calculated over the entire set of programming problems. The parameter $n$ indicates the number of generated code snippets for each programming problem, and the variable $c$ denotes the count of correct code snippets that successfully pass all the test cases. A lower \textit{Pass@k} indicates that the model is less accurate in solving the programming problems, and thus these programming problems are less likely to overlap with the models' training dataset.
Consistent with previous research, we set $k = 1, 10, 100$ for our evaluation.
Additionally, we use the relative \textit{Pass@k} drop as another evaluation metric. It allows us to directly compare the performance of the original and newly generated programming problems. This metric is calculated as the percentage difference in \textit{Pass@k} values between the two sets. 

%% file: Setup/Metric/natural.tex
\fakeparagraph{Naturalness Metrics}
To measure the naturalness of the generated programming problems, we utilize three metrics.
Our first metric is \textit{Perplexity}, which is widely used to measure the naturalness of natural language sentences. \textit{Perplexity} is a metric that evaluates how well a language model predicts a sequence of words.
Formally, given a tokenized problem description as $(x_1, x_2, \cdots, x_t)$, \textit{perplexity} is computed as follows:
\begin{equation}
\label{eq:perplexity}
\text{Perplexity} = \text{EXP} \left\{-\frac{1}{t} \sum_{i=1}^{t} \log p(x_i ; | x_1, \cdots, x_{i-1}) \right\}
\end{equation}
Here, $p(x_i ; | x_1, \cdots, x_{i-1})$ represents the probabilities obtained using a well-trained language model to predict the $i^{th}$ token. A well-trained language model is expected to assign higher probabilities to more natural and fluent sentences, while less natural or grammatically incorrect sentences will receive lower probabilities.
A lower \textit{perplexity} value indicates that the language model is more confident and accurate in predicting the sentence, thereby suggesting that the sentence is more natural and fluent. 
Our second metric is the \textit{Number of IDE Warnings}, which measures the total number of warnings of the generated programming problem prompts. 
A lower number of IDE warnings signifies that the programming prompts are more natural and realistic.
Our final metric for assessing naturalness is \textit{Human Scores}, wherein we enlist the help of volunteers to assign scores to each programming problem prompt. Following established practices \cite{recode}, we distribute each programming prompt to five human annotators who possess familiarity with Python. These annotators are tasked with rating the naturalness on a scale ranging from 0 (not natural) to 0.5 (possible but rare in practice) to 1 (natural).

%% file: Setup/Metric/stable.tex

%% file: Setup/process.tex
\vspace{-2mm}
\subsection{Experiment Process}

\input{Setup/Process/diversity}

\input{Setup/Process/effectiveness}

\input{Setup/Process/natural}

\input{Setup/Process/stable}

%% file: Setup/Process/diversity.tex
\fakeparagraph{\ref{rq:diversity} Process} 
To tackle RQ1, we perform two series of experiments aimed at evaluating the \textit{external diversity} and \textit{internal diversity} of the generated programming problems, respectively.
External diversity quantifies the dissimilarity between the newly generated programming problems and the original seed programming problems.
Conversely, internal diversity measures the variability within each problem-generation method across multiple trials.
In \textit{external diversity} evaluation, each programming problem within our evaluation datasets serves as a seed problem. We then apply each problem-generation method to create new programming problems and measure diversity metrics by comparing the seed problem to the generated one.
For our \textit{internal diversity} evaluation, we conducted two experiments.
In our first experiment, we conducted two runs for each method. As a result, for each seed problem, we obtain two versions of the new problem generated by each method. We compute the diversity metric by comparing these two versions of the problems.
In our second experiment, our goal is to showcase the effective ability of \tool in ensuring long-term data integrity. 
To achieve this, we run \tool once on the seed problem, resulting in the creation of the initial version of the new programming problems. Following this, we iteratively run \tool $K$ times, generating $K$ distinct versions. 
After that, we calculate the different implementation rates (\textit{DiffImp}) by comparing the initial generated version with the $K$ subsequent versions. 
If a problem in the initial version does not match any implementation in the $K$ versions, we consider it to have a different implementation. Otherwise, we treat it as the same implementation.
To ensure statistical robustness and mitigate the influence of randomness, we repeat each experiment ten times and report the average values of the metrics.



%% file: Setup/Process/effectiveness.tex
\fakeparagraph{\ref{rq:effectiveness} Process} 
To address RQ2, we conducted two experiments.
In our first experiment, we aim to explore the effectiveness of our generated semantic-diverse programming problems in uncovering the limitations of existing \lcgms. To achieve this objective, we utilize all problem-generation methods to create programming problems. Subsequently, we input each problem's prompt into each \lcgm and collect the generated solution programs. Following this, we execute the canonical solution on the test inputs to obtain the expected outputs. Finally, we execute these generated solution programs on the provided test inputs and collect the resulting outputs. Ultimately, we compare the executed outputs with the expected outputs to analyze the effectiveness of the methods. If our generated semantic-diverse programming problems are effective in uncovering the limitations of existing \lcgms, then each \lcgm would show a significant drop on \textit{Pass@k} metric.
For our second experiment, our goal is to investigate whether the drop in correctness of \lcgm is primarily due to inherent limitations rather than the level of challenge presented by our \textit{lambda programming problem}. To do this, we feed our \textit{lambda programming problem} to each \lcgm and measure the \textit{Pass@k} metric with the generated programs. If each \lcgm can produce high \textit{Pass@k} on our \textit{lambda programming problem}, then the effectiveness drop in our first experiment can be attributed to the inherent limitations of \lcgm rather than the complexity of our \textit{lambda programming problem}.

%% file: Setup/Process/natural.tex
\fakeparagraph{\ref{rq:naturalness} Process}  
We begin by performing a quantitative evaluation to measure the \textit{perplexity} of the problem prompts generated by each method. Calculating the \textit{perplexity} of these prompts (\equref{eq:perplexity}) requires a well-trained language model to predict each token probability in the prompts. We follow established methodology \cite{gpt2} and employ \texttt{GPT-2} as our language model due to it being trained on large-scale natural language corpus and widely used in existing work for measuring \textit{perplexity}.
Later, we copy all the generated programming problem prompts into a widely used Python Integrated Development Environment (IDE), namely \textit{PyCharm}. We then count the IDE warnings of each method, considering that the original programming problem prompts contained some pre-existing warnings stemming from typos errors. Hence, when calculating the number of IDE warnings, we exclude those that were part of the original dataset.
Finally, we conduct a human study to investigate the naturalness of the generated programming problem prompt. Specifically, we follow existing work \cite{wang2022recode} and conduct a human study.
In the human study, we assign each problem description to five human annotators who possess familiarity with the Python language. These annotators are asked to rate the naturalness of the problems on a scale from 0 to 1. Specifically, the scale is defined as follows: 0 denotes \textit{not natural}, 0.5 indicates \textit{possible to appear in practice but rare}, and 1 represents \textit{natural}.
For the evaluation of naturalness, we exclude demo modification methods (\eg \texttt{Add Demo}, \texttt{Remove Demo}, and \texttt{Replace Demo}) as they do not modify the problem description.

%% file: Setup/Process/stable.tex
\fakeparagraph{\ref{rq:sensitivity} Process}
To showcase the stability of benchmarking results and consistent performance across varied hyperparameters by \tool, we conduct two series of experiments.
In our first experiment, we aim to illustrate that despite the introduction of randomness in the programming problem generation process, this randomness does not significantly impact benchmarking results. To achieve this objective, we run \tool five trials and utilize the programming problems generated in each trial to benchmark every \lcgm. Subsequently, we present the averaged \textit{Pass@k} and its variance.
In the second series of experiments, our aim is to demonstrate the robustness of \tool concerning hyperparameter configurations in \lcgms. Specifically, we conduct two experiments to explore the impact of different \lcgm inference settings on the performance of \tool. Our focus is on analyzing \tool's responsiveness to two critical hyperparameters: the number of code candidates (denoted as $n$ in \equref{eq:pass}) and sampling temperatures.
Regarding the hyperparameter concerning the number of code candidates, we set $n$ within a range spanning from 10 to 100 and evaluate the efficacy metric \textit{Pass@1}.
Regarding the temperature hyperparameter, we vary the temperature across the range of 0.1 to 0.9, and we present the \textit{Pass@1} and \textit{Pass@10} metrics of \tool.
Due to limitations in space, we present results for the \textit{HumanEval} dataset with four models exclusively.

%% file: Setup/implementation.tex
\vspace{-2mm}
\subsection{Implementation Details}

We conducted our evaluation on a server equipped with an Intel Xeon E5-26 CPU and eight NVIDIA A4500 GPUs.
For each problem, we generated 100 candidate solutions per model and computed the effectiveness metric.
We then examined the impact of the number of candidates in \secref{sec:sensitive} to gain insights into how it affects performance.
In using \lcgms to generate code solutions, we set the inference temperature as 0.7, following existing literature, and we further explored the effect of temperature in \secref{sec:sensitive}.
To ensure the reliability of our findings for each research question, we ran each method multiple times and reported averaged results, mitigating randomness influence.

%% file: Text/evaluation.tex
\section{Evaluation Results}

\input{Evaluation/rq2}

\input{Evaluation/rq1}

\input{Evaluation/rq3}

\input{Evaluation/rq4}

%% file: Evaluation/rq2.tex
\vspace{-1mm}
\subsection{\ref{rq:diversity} Diversity}

\input{Table/rq2}

\fakeparagraph{External Diversity Results} 
The results of external diversity are presented in the left section of \tabref{tab:diversity}.
From the results, we have the following observations: 
Firstly, regarding prompt diversity, both linguistic and semantic aspects of the prompts exhibit lower diversity in our approaches compared to the baseline methods. This indicates that our approach excels in generating diverse problem descriptions, primarily by proposing a concrete and novel programming problem description distinct from the original, whereas the baseline methods merely modify certain tokens in the descriptions.
Secondly, in the context of solutions, \tool achieves a 100\% different implementation rate, while all baseline methods consistently yield a rate of 0. This stems from our innovative programming problem merging technique, allowing \tool to generate varied canonical solutions unlike the original canonical solutions, a capability lacking in the baseline methods.


\fakeparagraph{Internal Diversity Results} 
The internal diversity results are presented in the right section of \tabref{tab:diversity}.
From the results, we note that \texttt{Token Mutation} achieves a higher level of diversity in the prompt part. This is attributed to the random token mutations performed in each run by \texttt{Token Mutation}. On the other hand, \tool modifies only the \textit{offset} value within its template, resulting in lower prompt diversity when comparing two different trial runs.
However, \tool demonstrates a notable advantage in solution diversity. Specifically, in our \textit{pure value transformation}, we observed considerably lower solution diversity compared to our \textit{type-aware value transformation}. This disparity arises from certain programming problems in the original dataset that only return \textit{boolean} type values. Without altering the return value type, the random search space for \textit{boolean} values remains limited to two possibilities, such as \textit{True} and \textit{False}.

\begin{figure}[tbp]
 \begin{minipage}[t]{0.34\textwidth}
    \centering
    \includegraphics[width=0.99\textwidth]{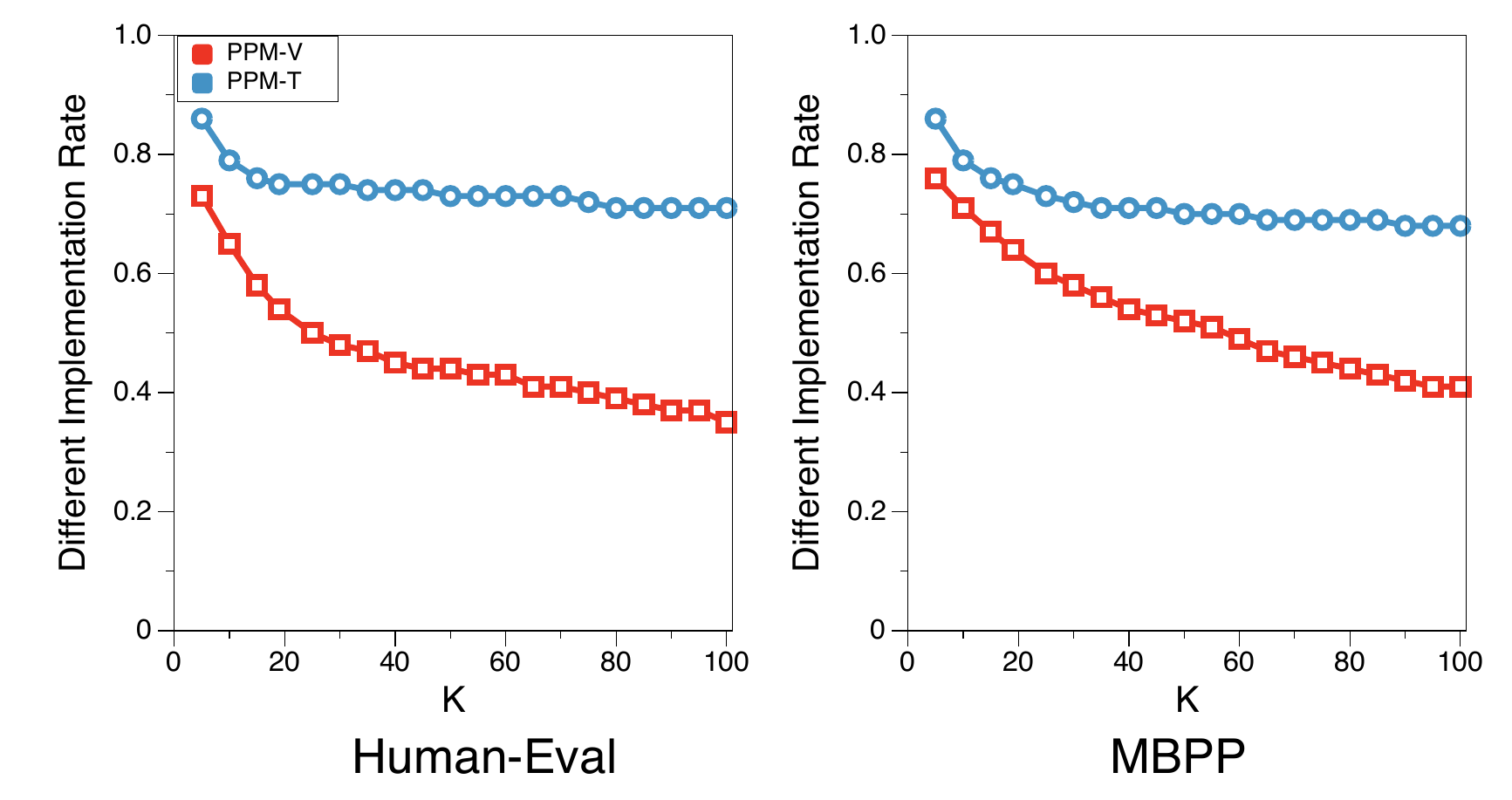}
    \vspace{-8mm}
    \caption{\textit{DiffImp} of \tool}
    \label{fig:curve}
  \end{minipage}%
  \begin{minipage}[t]{0.63\textwidth}
    \centering
    \includegraphics[width=0.99\textwidth]{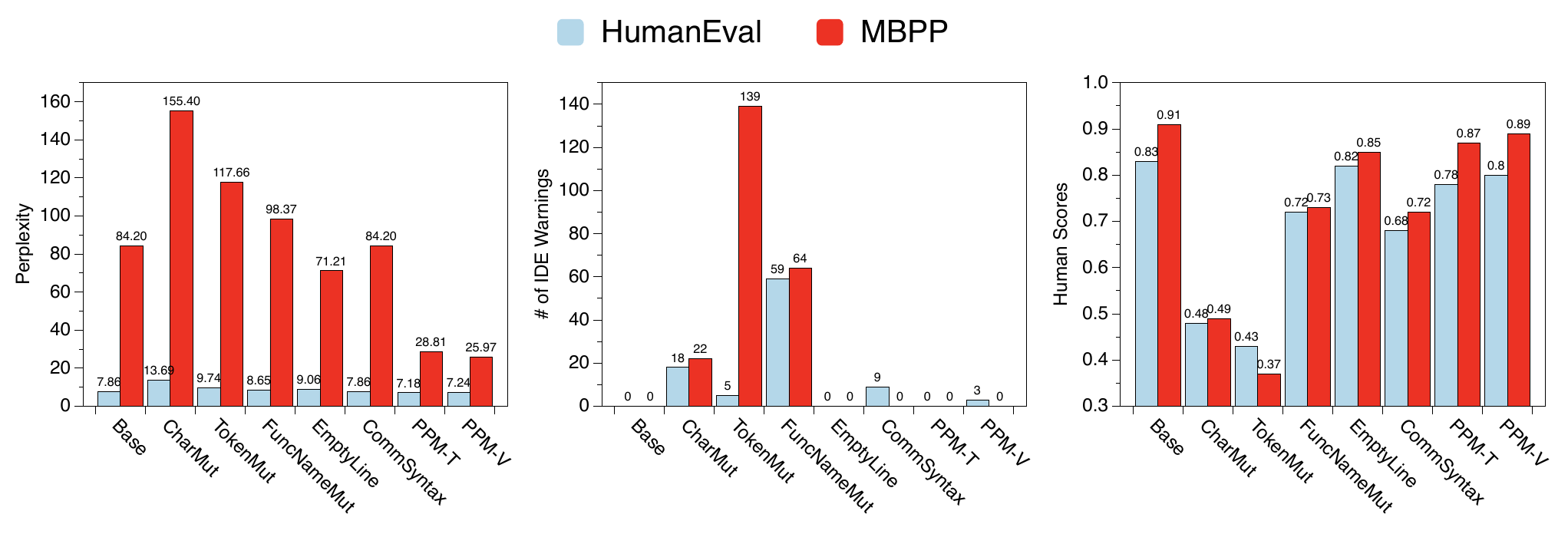}
    \vspace{-6mm}
    \caption{Naturalness results}
    \label{fig:natural}
  \end{minipage}
\end{figure}

The different implementation rates achieved by \tool through multiple trials are depicted in \figref{fig:curve}. 
Examining the outcomes, we observe that as $k$ increases from 0 to 100, the programming problems exhibiting different semantics (i.e., requiring distinct implementations) decrease initially and then stabilize. These findings indicate that even after 100 attempts, approximately 40\% of the programming problems for \texttt{PPM-V} and 70\% for \texttt{PPM-T} remain unrepeated.
These results imply the efficacy of \tool in maintaining long-term data integrity and its resilience against potential risks of training data leakage, even when a version of programming problems from \tool is publicly available on the Internet.

\begin{center}
\begin{tcolorbox}[colback=blue!8,
                  colframe=black,
                  width=0.99\textwidth,
                  arc=1mm, auto outer arc,
                  boxrule=0.9pt,
                 ]
\small Answers to \textbf{\ref{rq:diversity}}: Based on our experimental findings, it is evident that \tool can generate programming problems that deviate from the initial seed, presenting diverse problem descriptions and solutions. Moreover, upon executing \tool multiple times, the likelihood of generating programming problems that share identical solutions is minimal. \normalsize
\end{tcolorbox}
\end{center}

%% file: Table/rq2.tex
\begin{table}[tbp]
  \centering
  \caption{Diversity Results}
  \vspace{-3mm}
  \resizebox{0.78\textwidth}{!}{
    \begin{NiceTabular}{l|cccccc|cccccc}
    \CodeBefore
        \rowcolors{2}{}{blue!8}
        \Body
    \toprule
    \toprule
    \multicolumn{1}{c|}{\multirow{3}[2]{*}{\textbf{Methods}}} & \multicolumn{6}{c|}{\textbf{External Diversity}} & \multicolumn{6}{c}{\textbf{External Diversity}} \\
          & \multicolumn{3}{c}{\textbf{HumanEval}} & \multicolumn{3}{c|}{\textbf{MBPP}} & \multicolumn{3}{c}{\textbf{HumanEval}} & \multicolumn{3}{c}{\textbf{MBPP}} \\
          & \textbf{BLEU-4 $\downarrow$} & \textbf{SemSim $\downarrow$} & \textbf{DiffImp $\uparrow$} & \textbf{BLEU-4 $\downarrow$} & \textbf{SemSim $\downarrow$} & \textbf{DiffImp $\uparrow$} & \textbf{BLEU-4 $\downarrow$} & \textbf{SemSim $\downarrow$} & \textbf{DiffImp $\uparrow$} & \textbf{BLEU-4 $\downarrow$} & \textbf{SemSim $\downarrow$} & \textbf{DiffImp $\uparrow$} \\
    \midrule
    \textbf{Base} & 1.00  & 1.00  & 0.00  & 1.00  & 1.00  & 0.00  & 1.00  & 1.00  & 0.00  & 1.00  & 1.00  & 0.00  \\
    \midrule
    \textbf{Add Demo} & 1.00  & 1.00  & 0.00  & 0.86  & 1.00  & 0.00  & 1.00  & 1.00  & 0.00  & 1.00  & 1.00  & 0.00  \\
    \textbf{Del Demo} & 1.00  & 1.00  & 0.00  & -     & -     & 0.00  & 1.00  & 1.00  & 0.00  & 1.00  & 1.00  & 0.00  \\
    \textbf{Rep Demo} & 1.00  & 1.00  & 0.00  & -     & -     & 0.00  & 1.00  & 1.00  & 0.00  & 1.00  & 1.00  & 0.00  \\
    \textbf{Token Mutation} & 0.82  & 0.96  & 0.00  & 0.76  & 0.95  & 0.00  & \textbf{0.72 } & \textbf{0.95 } & 0.00  & \textbf{0.66 } & \textbf{0.92 } & 0.00  \\
    \textbf{Char Mutation} & 0.84  & 0.97  & 0.00  & 0.78  & 0.92  & 0.00  & 0.81  & 0.97  & 0.00  & 0.78  & 0.94  & 0.00  \\
    \textbf{Func Mutation} & 0.98  & 1.00  & 0.00  & 0.98  & 1.00  & 0.00  & 1.00  & 1.00  & 0.00  & 1.00  & 1.00  & 0.00  \\
    \textbf{Insert Line} & 1.00  & 1.00  & 0.00  & 1.00  & 1.00  & 0.00  & 1.00  & 1.00  & 0.00  & 1.00  & 1.00  & 0.00  \\
    \textbf{CommSyntax} & 0.81  & 0.98  & 0.00  & 0.73  & 0.99  & 0.00  & 1.00  & 1.00  & 0.00  & 1.00  & 1.00  & 0.00  \\
    \midrule
    \textbf{PPM-V} & 0.69  & \textbf{0.89 } & \textbf{1.00 } & 0.57  & \textbf{0.84 } & \textbf{1.00 } & 0.97  & 0.96  & 0.75  & 0.96  & 0.94  & 0.76  \\
    \textbf{PPM-T} & \textbf{0.66 } & 0.90  & \textbf{1.00 } & \textbf{0.54 } & 0.90  & \textbf{1.00 } & 0.84  & 0.97  & \textbf{0.98 } & 0.81  & 0.96  & \textbf{0.97 } \\
    \bottomrule
    \bottomrule
    \end{NiceTabular}%
    }
  \label{tab:diversity}%
\end{table}%

%% file: Evaluation/rq1.tex
\vspace{-2mm}
\subsection{\ref{rq:effectiveness}: Effectiveness}

\input{Table/rq1_human}

\input{Table/rq1_mbpp}

\fakeparagraph{New Problem Results} The effectiveness results are presented in \tabref{tab:rq1_humaneval} and \tabref{tab:rq1_mbpp}, where \tabref{tab:rq1_humaneval} corresponds to the \textit{HumanEval} dataset, and \tabref{tab:rq1_mbpp} corresponds to the \textit{MBPP} dataset.
As the \textit{MBPP} dataset is a zero-shot dataset with no demonstrations in its prompts, the evaluation methods \texttt{Del Demo} and \texttt{Replace Demo} cannot be applied to this dataset.
The numbers displayed without brackets in the tables represent the measured \textit{Pass@k} values, reflecting the functional correctness of the generated code. On the other hand, the numbers enclosed in brackets represent the relative \textit{Pass@k} drop compared with the \texttt{Base} model. This drop quantifies the change in performance between the original programming problems and the newly crafted programming problems.

From the results in \tabref{tab:rq1_humaneval} and \tabref{tab:rq1_mbpp}, we observe:
\textit{(1)} In all settings, \tool demonstrates a remarkable capability to significantly decrease the code generation model's performance, setting it apart from the baseline. 
For example, the \textit{Pass@1} value decreased by 90\% and 85\% for our approaches. In contrast, the \textit{Pass@1} values either remained the same or slightly dropped by about 5\% to 15\% for the baseline methods. 
This is because our newly crafted programming problem changes the semantics of the problem description, while the baseline approaches do not. 
\textit{(2)} The baseline method can even increase the model's functional correctness. For instance, the \textit{Pass@1} values increase by 16.77\% when deleting or replacing input/output demos in the prompts. 
\textit{(3)} More demos do not always yield better correctness. For instance, when $k$ is set at 10 and 100, the \textit{Pass@k} values for the \textsf{Incoder-1b} drop by about 6.25\% and 21.43\%, respectively, in the \textit{HumanEval} dataset. 



\input{Table/lambda}

\fakeparagraph{Lambda Programming Problem Results}
The \textit{Pass@k} scores for each \lcgm in our \textit{lambda programming problem} are presented in \tabref{tab:lambda}.
For each type of \textit{lambda programming problem}, each \lcgm achieves almost perfect accuracy.
Upon comparing these scores with the \textit{Pass@1} values from \tabref{tab:rq1_humaneval} and \tabref{tab:rq1_mbpp}, a notable observation emerges: the \textit{Pass@k} scores for our \textit{lambda programming problem} significantly surpass those of the original dataset. These results imply that our proposed \textit{lambda programming problem} is not considerably more challenging for \lcgm to understand. 
Consequently, the reduction in \lcgm's \textit{Pass@k} for the merged programming problem can be attributed to the merging concept, rather than our specific \textit{lambda programming problem}.


\begin{center}
\begin{tcolorbox}[colback=blue!8,
                  colframe=black,
                  width=0.98\textwidth,
                  arc=1mm, auto outer arc,
                  boxrule=0.9pt,
                 ]
\small Answers to \textbf{\ref{rq:effectiveness}}: \tool can generate  problems that reduce code generation model performance by modifying the problem semantics, a feature not found in existing methods. \normalsize

\end{tcolorbox}
\end{center}

%% file: Table/rq1_human.tex
\begin{table*}[t!]
    \centering
    \caption{The effectiveness evaluation on the HumanEval dataset.}
    \vspace{-3mm}
    \label{tab:rq1_humaneval}
    \resizebox{\linewidth}{!}{
        \begin{NiceTabular}{lcccccccccccc}
        \CodeBefore
            \rowcolors{2}{}{blue!8}
        \Body
    \toprule
    \toprule
    \textbf{Methods} & \multicolumn{3}{c}{\textbf{CodeGen-2b}} & \multicolumn{3}{c}{\textbf{CodeGen-6b}} & \multicolumn{3}{c}{\textbf{CodeGen2-3.7b}} & \multicolumn{3}{c}{\textbf{CodeGen2-1b}} \\
\cmidrule{2-13}          & \textbf{Pass@1} & \textbf{Pass@10} & \textbf{Pass@100} & \textbf{Pass@1} & \textbf{Pass@10} & \textbf{Pass@100} & \textbf{Pass@1} & \textbf{Pass@10} & \textbf{Pass@100} & \textbf{Pass@1} & \textbf{Pass@10} & \textbf{Pass@100} \\
    \midrule
    \textbf{Base} & 0.2   & 0.41  & 0.6   & 0.24  & 0.48  & 0.69  & 0.11  & 0.23  & 0.37  & 0.08  & 0.15  & 0.25 \\
    \midrule
    \textbf{Add Demo} & 0.2 (0.00\%) & 0.41 (0.00\%) & 0.62 (+3.33\%) & 0.24 (0.00\%) & 0.49 (+2.08\%) & 0.75 (+8.70\%) & 0.11 (0.00\%) & 0.24 (+4.35\%) & 0.4 (+8.11\%) & 0.07 (-12.50\%) & 0.15 (0.00\%) & 0.29 (+16.00\%) \\
    \textbf{Del Demo} & 0.2 (0.00\%) & 0.41 (0.00\%) & 0.64 (+6.67\%) & 0.24 (0.00\%) & 0.49 (+2.08\%) & 0.7 (+1.45\%) & 0.11 (0.00\%) & 0.22 (-4.35\%) & 0.37 (0.00\%) & 0.08 (0.00\%) & 0.15 (0.00\%) & 0.25 (0.00\%) \\
    \textbf{Rep Demo} & 0.2 (0.00\%) & 0.4 (-2.44\%) & 0.61 (+1.67\%) & 0.24 (0.00\%) & 0.48 (0.00\%) & 0.73 (+5.80\%) & 0.11 (0.00\%) & 0.24 (+4.35\%) & 0.39 (+5.41\%) & 0.08 (0.00\%) & 0.15 (0.00\%) & 0.26 (+4.00\%) \\
    \textbf{Token Mutation} & 0.19 (-5.00\%) & 0.4 (-2.44\%) & 0.58 (-3.33\%) & 0.22 (-8.33\%) & 0.44 (-8.33\%) & 0.67 (-2.90\%) & 0.1 (-9.09\%) & 0.22 (-4.35\%) & 0.35 (-5.41\%) & 0.06 (-25.00\%) & 0.14 (-6.67\%) & 0.26 (+4.00\%) \\
    \textbf{Char Mutation} & 0.17 (-15.00\%) & 0.36 (-12.20\%) & 0.53 (-11.67\%) & 0.22 (-8.33\%) & 0.44 (-8.33\%) & 0.63 (-8.70\%) & 0.1 (-9.09\%) & 0.23 (0.00\%) & 0.39 (+5.41\%) & 0.07 (-12.50\%) & 0.13 (-13.33\%) & 0.21 (-16.00\%) \\
    \textbf{FuncName Mutation} & 0.19 (-5.00\%) & 0.4 (-2.44\%) & 0.59 (-1.67\%) & 0.23 (-4.17\%) & 0.47 (-2.08\%) & 0.72 (+4.35\%) & 0.1 (-9.09\%) & 0.23 (0.00\%) & 0.39 (+5.41\%) & 0.06 (-25.00\%) & 0.14 (-93.33\%) & 0.24 (-4.00\%) \\
    \textbf{Insert Line} & 0.19 (-5.00\%) & 0.41 (0.00\%) & 0.67 (+11.67\%) & 0.24 (0.00\%) & 0.48 (0.00\%) & 0.7 (+1.45\%) & 0.11 (0.00\%) & 0.25 (+8.70\%) & 0.39 (+5.41\%) & 0.08 (0.00\%) & 0.14 (-6.67\%) & 0.24 (-4.00\%) \\
    \textbf{CommSyntax} & 0.17 (-15.00\%) & 0.37 (-9.76\%) & 0.63 (+5.00\%) & 0.19 (-20.83\%) & 0.43 (-10.42\%) & 0.7 (+1.45\%) & 0.07 (-36.36\%) & 0.19 (-17.39\%) & 0.34 (-8.11\%) & 0.06 (-25.00\%) & 0.13 (-13.33\%) & 0.22 (-12.00\%) \\
    \midrule
    \textbf{PPM-V} & \textbf{0.02 (-90.00\%)} & \textbf{0.1 (-75.61\%)} & 0.25 (-58.33\%) & \textbf{0.03 (-87.50\%)} & 0.12 (-75.00\%) & 0.26 (-62.32\%) & \textbf{0 (-99.99\%)} & \textbf{0.03 (-86.96\%)} & \textbf{0.09 (-75.68\%)} & \textbf{0 (-99.99\%)} & \textbf{0.02 (-86.67\%)} & \textbf{0.08 (-68.00\%)} \\
    \textbf{PPM-T} & 0.01 (-95.00\%) & 0.07 (-82.93\%) & \textbf{0.16 (-73.33\%)} & 0.02 (-91.67\%) & \textbf{0.09 (-81.25\%)} & \textbf{0.21 (-69.57\%)} & 0.01 (-90.91\%) & 0.03 (-86.96\%) & 0.08 (-78.38\%) & 0 (-99.99\%) & 0.01 (-93.33\%) & 0.06 (-76.00\%) \\
    \midrule
    \midrule
    \textbf{Methods} & \multicolumn{3}{c}{\textbf{Incoder-1b}} & \multicolumn{3}{c}{\textbf{Incoder-6b}} & \multicolumn{3}{c}{\textbf{Santacoder-1.1b}} & \multicolumn{3}{c}{\textbf{PolyCoder}} \\
\cmidrule{2-13}          & \textbf{Pass@1} & \textbf{Pass@10} & \textbf{Pass@100} & \textbf{Pass@1} & \textbf{Pass@10} & \textbf{Pass@100} & \textbf{Pass@1} & \textbf{Pass@10} & \textbf{Pass@100} & \textbf{Pass@1} & \textbf{Pass@10} & \textbf{Pass@100} \\
    \midrule
    \textbf{Base} & 0.06  & 0.16  & 0.28  & 0.11  & 0.28  & 0.48  & 0.15  & 0.29  & 0.45  & 0.04  & 0.1   & 0.21 \\
    \midrule
    \textbf{Add Demo} & 0.06 (0.00\%) & 0.15 (-6.25\%) & 0.22 (-21.43\%) & 0.11 (0.00\%) & 0.27 (-3.57\%) & 0.42 (-12.50\%) & 0.15 (0.00\%) & 0.3 (+3.45\%) & 0.47 (+4.44\%) & 0.04 (0.00\%) & 0.1 (0.00\%) & 0.17 (-19.05\%) \\
    \textbf{Del Demo} & 0.07 (+16.67\%) & 0.16 (0.00\%) & 0.27 (-3.57\%) & 0.1 (-9.09\%) & 0.26 (-7.14\%) & 0.44 (-8.33\%) & 0.15 (0.00\%) & 0.3 (+3.45\%) & 0.48 (+6.67\%) & 0.04 (0.00\%) & 0.1 (0.00\%) & 0.18 (-14.29\%) \\
    \textbf{Rep Demo} & 0.07 (+16.67\%) & 0.17 (+6.25\%) & 0.31 (+10.71\%) & 0.11 (0.00\%) & 0.27 (-3.57\%) & 0.45 (-6.25\%) & 0.15 (0.00\%) & 0.3 (+3.45\%) & 0.49 (+8.89\%) & 0.04 (0.00\%) & 0.1 (0.00\%) & 0.16 (-23.81\%) \\
    \textbf{Token Mutation} & 0.05 (-16.67\%) & 0.15 (-6.25\%) & 0.24 (-14.29\%) & 0.1 (-9.09\%) & 0.25 (-10.71\%) & 0.45 (-6.25\%) & 0.14 (-6.67\%) & 0.28 (-3.45\%) & 0.47 (+4.44\%) & 0.03 (-25.00\%) & 0.1 (0.00\%) & 0.19 (-9.52\%) \\
    \textbf{Char Mutation} & 0.06 (0.00\%) & 0.15 (-6.25\%) & 0.29 (+3.57\%) & 0.09 (-18.18\%) & 0.24 (-14.29\%) & 0.42 (-12.50\%) & 0.13 (-13.33\%) & 0.28 (-3.45\%) & 0.45 (0.00\%) & 0.03 (-25.00\%) & 0.09 (-10.00\%) & 0.18 (-14.29\%) \\
    \textbf{FuncName Mutation} & 0.06 (0.00\%) & 0.15 (-6.25\%) & 0.26 (-7.14\%) & 0.1 (-9.09\%) & 0.26 (-7.14\%) & 0.43 (-10.42\%) & 0.14 (-6.67\%) & 0.28 (-3.45\%) & 0.48 (+6.67\%) & 0.04 (0.00\%) & 0.1 (0.00\%) & 0.20 (-4.76\%) \\
    \textbf{Insert Line} & 0.07 (+16.67\%) & 0.16 (0.00\%) & 0.27 (-3.57\%) & 0.11 (0.00\%) & 0.26 (-7.14\%) & 0.45 (-6.25\%) & 0.15 (0.00\%) & 0.3 (+3.45\%) & 0.49 (+8.89\%) & 0.04 (0.00\%) & 0.09 (-10.00\%) & 0.19 (-9.52\%) \\
    \textbf{CommSyntax} & 0.04 (-33.33\%) & 0.13 (-18.75\%) & 0.23 (-17.86\%) & 0.09 (-18.18\%) & 0.25 (-10.71\%) & 0.42 (-12.50\%) & 0.12 (-20.00\%) & 0.28 (-3.45\%) & 0.44 (-2.22\%) & 0.03 (-25.00\%) & 0.09 (-10.00\%) & 0.18 (-14.29\%) \\
    \midrule
    \textbf{PPM-V} & \textbf{0.01 (-83.33\%)} & 0.04 (-75.00\%) & 0.12 (-57.14\%) & 0.02 (-81.82\%) & 0.08 (-71.43\%) & 0.16 (-66.67\%) & \textbf{0.02 (-86.67\%)} & \textbf{0.07 (-75.86\%)} & 0.16 (-64.44\%) & 0 (-99.99\%) & 0.01 (-90.00\%) & 0.08 (-61.90\%) \\
    \textbf{PPM-T} & 0 (-99.99\%) & \textbf{0.02 (-87.50\%)} & \textbf{0.07 (-75.00\%)} & \textbf{0.01 (-90.91\%)} & \textbf{0.04 (-85.71\%)} & \textbf{0.11 (-77.08\%)} & 0.01 (-93.33\%) & 0.04 (-86.21\%) & \textbf{0.12 (-73.33\%)} & 0.01 (-75.00\%) & 0.01 (-90.00\%) & 0.06 (-71.43\%) \\
    \bottomrule
    \bottomrule
    \end{NiceTabular}
    }
\end{table*}

%% file: Table/rq1_mbpp.tex
\begin{table*}[t!]
    \centering
    \caption{The effectiveness evaluation on the MBPP dataset.}
    \vspace{-4mm}
    \label{tab:rq1_mbpp}
    \resizebox{\linewidth}{!}{%
    \begin{NiceTabular}{lcccccccccccc}
        \CodeBefore
            \rowcolors{2}{}{blue!8}
        \Body
    \toprule
    \toprule
    \textbf{Methods} & \multicolumn{3}{c}{\textbf{CodeGen-2b}} & \multicolumn{3}{c}{\textbf{CodeGen-6b}} & \multicolumn{3}{c}{\textbf{CodeGen2-3.7b}} & \multicolumn{3}{c}{\textbf{CodeGen2-1b}} \\
\cmidrule{2-13}          & \textbf{Pass@1} & \textbf{Pass@10} & \textbf{Pass@100} & \textbf{Pass@1} & \textbf{Pass@10} & \textbf{Pass@100} & \textbf{Pass@1} & \textbf{Pass@10} & \textbf{Pass@100} & \textbf{Pass@1} & \textbf{Pass@10} & \textbf{Pass@100} \\
    \midrule
    \textbf{Base} & 0.36  & 0.67  & 0.82  & 0.39  & 0.70  & 0.84  & 0.17  & 0.45  & 0.65  & 0.13  & 0.37  & 0.57  \\
    \midrule
    \textbf{Add Demo} & 0.37 (+2.78\%) & 0.68 (+1.49\%) & 0.83 (+1.22\%) & 0.45 (+15.38\%) & 0.73 (+4.29\%) & 0.86 (+2.38\%) & 0.22 (+29.41) & 0.52 (+15.56\%) & 0.71 (+9.23\%) & 0.13 (0.00\%) & 0.39 (+5.41\%) & 0.61 (+7.02\%) \\
    \textbf{Token Mutation} & 0.32 (-11.11\%) & 0.63 (-5.97\%) & 0.79 (-3.66\%) & 0.37 (-5.13\%) & 0.67 (-4.29\%) & 0.83 (-1.19\%) & 0.16 (-5.88\%) & 0.43 (-4.44\%) & 0.62 (-4.62\%) & 0.11 (-15.38\%) & 0.35 (-5.41\%) & 0.55 (-3.51\%) \\
    \textbf{Char Mutation} & 0.26 (-27.78\%) & 0.56 (-16.42\%) & 0.76 (-7.32\%) & 0.3 (-23.08\%) & 0.61 (-12.86\%) & 0.78 (-7.14\%) & 0.13 (-23.53\%) & 0.37 (-17.78\%) & 0.57 (-12.31\%) & 0.09 (-30.77\%) & 0.32 (-13.51\%) & 0.51 (-10.53\%) \\
    \textbf{FuncName Mutation} & 0.34 (-5.56\%) & 0.66 (-1.49\%) & 0.81 (-1.22\%) & 0.39 (0.00\%) & 0.7 (0.00\%) & 0.85 (+1.19\%) & 0.17 (0.00\%) & 0.45 (0.00\%) & 0.65 (0.00\%) & 0.12 (-7.69\%) & 0.37 (0.00\%) & 0.57 (0.00\%) \\
    \textbf{Insert Line} & 0.35 (-2.78\%) & 0.65 (-2.99\%) & 0.82 (0.00\%) & 0.39 (0.00\%) & 0.7 (0.00\%) & 0.85 (+1.19\%) & 0.16 (-5.88\%) & 0.44 (-2.22\%) & 0.63 (-3.08\%) & 0.12 (-7.69\%) & 0.35 (-5.41\%) & 0.56 (-1.75\%) \\
    \textbf{CommSyntax} & 0.26 (-27.78\%) & 0.61 (-8.96\%) & 0.8 (-2.44\%) & 0.29 (-25.64\%) & 0.65 (-7.14\%) & 0.83 (-1.19\%) & 0.12 (-29.41\%) & 0.39 (-13.33\%) & 0.57 (-12.31\%) & 0.08 (-38.46\%) & 0.3 (-18.92\%) & 0.53 (-7.02\%) \\
    \midrule
    \textbf{PPM-V} & 0.04 (-88.89\%) & 0.18 (-73.13\%) & 0.37 (-54.88\%) & 0.04 (-89.74\%) & 0.18 (-74.29\%) & 0.37 (-55.95\%) & 0.01 (-94.12\%) & 0.07 (-84.44\%) & 0.18 (-72.31\%) & 0.01 (-92.31\%) & 0.07 (-81.08\%) & 0.21 (-63.16\%) \\
    \textbf{PPM-T} & 0.06 (-83.33\%) & 0.22 (-67.16\%) & 0.39 (-52.44\%) & 0.06 (-84.62\%) & 0.21 (-70.00\%) & 0.4 (-52.38\%) & 0.03 (-82.35\%) & 0.12 (-73.33\%) & 0.24 (-63.08\%) & 0.02 (-84.62\%) & 0.11 (-70.27\%) & 0.25 (-56.14\%) \\
    \midrule
    \midrule
    \textbf{Methods} & \multicolumn{3}{c}{\textbf{Incoder-1b}} & \multicolumn{3}{c}{\textbf{Incoder-6b}} & \multicolumn{3}{c}{\textbf{Santacoder-1.1b}} & \multicolumn{3}{c}{\textbf{PolyCoder}} \\
\cmidrule{2-13}          & \textbf{Pass@1} & \textbf{Pass@10} & \textbf{Pass@100} & \textbf{Pass@1} & \textbf{Pass@10} & \textbf{Pass@100} & \textbf{Pass@1} & \textbf{Pass@10} & \textbf{Pass@100} & \textbf{Pass@1} & \textbf{Pass@10} & \textbf{Pass@100} \\
    \midrule
    \textbf{Base} & 0.12  & 0.39  & 0.59  & 0.13  & 0.44  & 0.65  & 0.26  & 0.56  & 0.71  & 0.08  & 0.30  & 0.52  \\
    \midrule
    \textbf{Add Demo} & 0.14 (+16.67\%) & 0.41 (+5.13\%) & 0.61 (+3.39\%) & 0.21 (+61.54\%) & 0.5 (+13.64\%) & 0.69 (+6.15\%) & 0.31 (+19.23\%) & 0.58 (+3.57\%) & 0.75 (5.63\%) & 0.10 (+25.00\%) & 0.32 (+6.67\%) & 0.53 (+1.92\%) \\
    \textbf{Token Mutation} & 0.1 (-16.67\%) & 0.37 (-5.13\%) & 0.56 (-5.08\%) & 0.12 (-7.69\%) & 0.41 (-6.82\%) & 0.64 (-1.54\%) & 0.24 (-7.69\%) & 0.54 (-3.57\%) & 0.7 (-1.41\%) & 0.06 (-25.00\%) & 0.28 (-6.67\%) & 0.51 (-1.92\%) \\
    \textbf{Char Mutation} & 0.09 (-25.00\%) & 0.33 (-15.38\%) & 0.52 (-11.86\%) & 0.1 (-23.08\%) & 0.36 (-18.18\%) & 0.6 (-7.69\%) & 0.19 (-26.92\%) & 0.47 (-16.07\%) & 0.68 (-4.23\%) & 0.07 (-12.50\%) & 0.25 (-16.67\%) & 0.44 (-15.38\%) \\
    \textbf{FuncName Mutation} & 0.12 (0.00\%) & 0.38 (-2.56\%) & 0.57 (-3.39\%) & 0.12 (-7.69\%) & 0.42 (-4.55\%) & 0.64 (-1.54\%) & 0.25 (-3.85\%) & 0.56 (0.00\%) & 0.73 (+2.82\%) & 0.08 (0.00\%) & 0.3 (0.00\%) & 0.5 (-3.85\%) \\
    \textbf{Insert Line} & 0.12 (0.00\%) & 0.39 (0.00\%) & 0.58 (-1.69\%) & 0.12 (-7.69\%) & 0.41 (-6.82\%) & 0.63 (-3.08\%) & 0.22 (-15.38\%) & 0.53 (-5.36\%) & 0.71 (0.00\%) & 0.07 (-12.5\%) & 0.28 (-6.67\%) & 0.51 (-1.92\%) \\
    \textbf{CommSyntax} & 0.05 (-58.33\%) & 0.23 (-41.03\%) & 0.48 (-18.64\%) & 0.09 (-30.77\%) & 0.36 (-18.18\%) & 0.6 (-7.69\%) & 0.23 (-11.54\%) & 0.55 (-1.79\%) & 0.74 (+4.23\%) & 0.08 (0.00\%) & 0.28 (-6.67\%) & 0.5 (-3.85\%) \\ \hline
    \textbf{PPM-V} & 0.01 (-91.67\%) & 0.08 (-79.49\%) & 0.23 (-61.02\%) & 0.02 (-84.62\%) & 0.11 (-75.00\%) & 0.27 (-58.46\%) & 0.03 (-88.46\%) & 0.14 (-75.00\%) & 0.32 (-54.93\%) & 0.01 (-87.50\%) & 0.07 (-76.67\%) & 0.18 (-65.38\%) \\
    \textbf{PPM-T} & 0.02 (-83.33\%) & 0.1 (-74.36\%) & 0.23 (-61.02\%) & 0.02 (-84.62\%) & 0.1 (-77.27\%) & 0.25 (-61.54\%) & 0.04 (-84.62\%) & 0.14 (-75.00\%) & 0.31 (-56.34\%) & 0 (-99.99\%) & 0.06 (-80.00\%) & 0.17 (-67.37\%) \\
    \bottomrule
    \bottomrule
    \end{NiceTabular}%
    }
\end{table*}

%% file: Table/lambda.tex
\begin{table}[tbp!]
  \centering
  \caption{The Pass@k of each \lcgm on our lambda programming problem}
    \vspace{-3mm}
    \resizebox{0.68\textwidth}{!}{
    \begin{NiceTabular}{c|cccccccccccc}
    \CodeBefore
            \rowcolors{1}{}{blue!8}
            \Body
    \toprule
    \toprule
    \multirow{6}[8]{*}{\textbf{PPT-V}} & \multicolumn{3}{c}{\textbf{CodeGen-2b}} & \multicolumn{3}{c}{\textbf{CodeGen-6b}} & \multicolumn{3}{c}{\textbf{CodeGen2-3.7b}} & \multicolumn{3}{c}{\textbf{CodeGen2-1b}} \\
\cmidrule{2-13}          & \textbf{Pass@1} & \textbf{Pass@10} & \textbf{Pass@100} & \textbf{Pass@1} & \textbf{Pass@10} & \textbf{Pass@100} & \textbf{Pass@1} & \textbf{Pass@10} & \textbf{Pass@100} & \textbf{Pass@1} & \textbf{Pass@10} & \textbf{Pass@100} \\
          & 0.51  & 0.91  & 0.92  & 0.50  & 0.89  & 0.92  & 0.13  & 0.59  & 0.75  & 0.14  & 0.56  & 0.83  \\
\cmidrule{2-13}          & \multicolumn{3}{c}{\textbf{Incoder-1b}} & \multicolumn{3}{c}{\textbf{Incoder-6b}} & \multicolumn{3}{c}{\textbf{Santacoder-1.1b}} & \multicolumn{3}{c}{\textbf{PolyCoder}} \\
\cmidrule{2-13}          & \textbf{Pass@1} & \textbf{Pass@10} & \textbf{Pass@100} & \textbf{Pass@1} & \textbf{Pass@10} & \textbf{Pass@100} & \textbf{Pass@1} & \textbf{Pass@10} & \textbf{Pass@100} & \textbf{Pass@1} & \textbf{Pass@10} & \textbf{Pass@100} \\
          & 0.16  & 0.55  & 0.83  & 0.42  & 0.87  & 0.92  & 0.48  & 0.88  & 0.92  & 0.06  & 0.36  & 0.67  \\
    \midrule
    \midrule
    \multirow{6}[8]{*}{\textbf{PPT-T}} & \multicolumn{3}{c}{\textbf{CodeGen-2b}} & \multicolumn{3}{c}{\textbf{CodeGen-6b}} & \multicolumn{3}{c}{\textbf{CodeGen2-3.7b}} & \multicolumn{3}{c}{\textbf{CodeGen2-1b}} \\
\cmidrule{2-13}          & \textbf{Pass@1} & \textbf{Pass@10} & \textbf{Pass@100} & \textbf{Pass@1} & \textbf{Pass@10} & \textbf{Pass@100} & \textbf{Pass@1} & \textbf{Pass@10} & \textbf{Pass@100} & \textbf{Pass@1} & \textbf{Pass@10} & \textbf{Pass@100} \\
          & 0.91  & 1.00  & 1.00  & 0.97  & 1.00  & 1.00  & 0.73  & 0.99  & 1.00  & 0.59  & 1.00  & 1.00  \\
\cmidrule{2-13}          & \multicolumn{3}{c}{\textbf{Incoder-1b}} & \multicolumn{3}{c}{\textbf{Incoder-6b}} & \multicolumn{3}{c}{\textbf{Santacoder-1.1b}} & \multicolumn{3}{c}{\textbf{PolyCoder}} \\
\cmidrule{2-13}          & \textbf{Pass@1} & \textbf{Pass@10} & \textbf{Pass@100} & \textbf{Pass@1} & \textbf{Pass@10} & \textbf{Pass@100} & \textbf{Pass@1} & \textbf{Pass@10} & \textbf{Pass@100} & \textbf{Pass@1} & \textbf{Pass@10} & \textbf{Pass@100} \\
          & 0.60  & 0.93  & 1.00  & 0.66  & 0.97  & 1.00  & 0.87  & 1.00  & 1.00  & 0.44  & 0.88  & 1.00  \\
    \bottomrule
    \bottomrule
    \end{NiceTabular}%
    }
  \label{tab:lambda}%
\end{table}%

%% file: Evaluation/rq3.tex
\vspace{-2mm}
\subsection{\ref{rq:naturalness}: Naturalness}


\fakeparagraph{Results}
The naturalness results are presented in \figref{fig:natural}. 
From left to right are the results of \textit{perplexity}, \textit{number of IDE warnings}, and \textit{human scores}.
The perplexity results demonstrate that our approaches outperform the baseline methods, as evidenced by their consistently lower perplexity values on both the \textit{HumanEval} dataset and the \textit{MBPP} dataset.
Interestingly, for the \textit{MBPP} dataset, our approaches even show an improvement in the naturalness of problem descriptions compared to the original descriptions. This can be attributed to the specific format of the problem descriptions in the \textit{MBPP} dataset, which follows the pattern of "Write a function to execute some command." Our added descriptions conform to the style of these commands, resulting in a more cohesive and natural overall description. The original short problem descriptions, when combined with our added descriptions, yield lower perplexity values, indicating an increase in naturalness and coherence. 
For the IDE warning metric, it is observed that \texttt{Token Mutation}, \texttt{Character Mutation}, and \texttt{FuncName Mutation} introduce a significant number of IDE warnings. 
This can be attributed to the inherent nature of these methods, which naturally introduce typos into the prompt. These typos are easily identified by the IDE, rendering these methods impractical and unrealistic.
We also observe three IDE warnings for \texttt{PPM-T} when applied to the \textit{Human-Eval} dataset. Upon manual examination of these warnings, we determined that they are because of a specific operator within \tool. This operator is designed to transform one string into another based on ASCII values. However, during the transformation of the string in the demo part of the prompt, the resulting output string may not be an English word. Consequently, the IDE detects these transformations as typos and thus these warnings are false positives.
For human score results. In line with the previous results, our approaches consistently achieved higher human scores compared to the baseline methods of token mutation and char mutation on both the \textit{HumanEval} dataset and the \textit{MBPP} dataset.
This is because \tool focus on generating natural and contextually relevant problem descriptions, resulting in higher human ratings for the quality and fluency of the generated descriptions.


\begin{center}
\begin{tcolorbox}[colback=blue!8,
                  colframe=black,
                  width=\textwidth,
                  arc=1mm, auto outer arc,
                  boxrule=0.9pt,
                 ]
\small Answers to \textbf{RQ3}: Based on our quantitative evaluation and human study, we conclude that \tool can generate natural and realistic programming problems. \normalsize

\end{tcolorbox}
\end{center}

%% file: Evaluation/rq4.tex
\subsection{\ref{rq:sensitivity}: Stability}
\label{sec:sensitive}

\begin{figure*}
    \centering
    \includegraphics[width=0.88\textwidth]{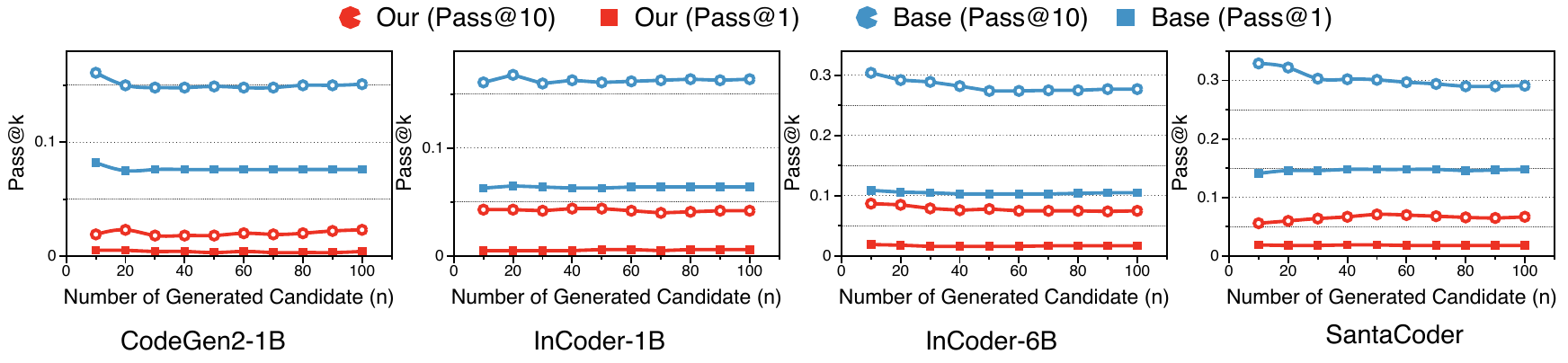}
    \vspace{-3mm}
    \caption{The effectiveness of \tool with different numbers of generated candidates (n)}
    \label{fig:candidate}
\end{figure*}

\input{Table/stability}

\fakeparagraph{Benchmarking Stability}
The stability results from multiple randomized trials of \tool are showcased in \tabref{tab:stability}. Each trial's \textit{Pass@k} values are presented alongside their corresponding averages and standard variances. Notably, all average values surpass three times the variance, implying a high degree of stability in the results. These results serve as confirmation that despite the introduction of randomness in the problem generation process, \tool consistently provides stable results.

\begin{figure*}
    \centering
    \includegraphics[width=0.88\textwidth]{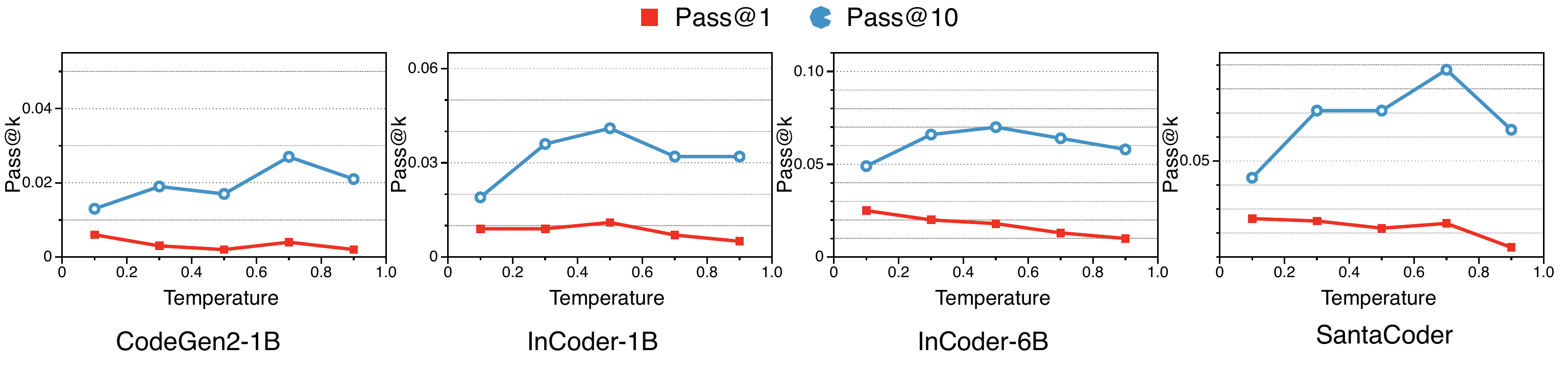}
    \vspace{-5mm}
    \caption{The effectiveness of \tool under different temperatures}
    \label{fig:temperature}
\end{figure*}

\fakeparagraph{Hyperparameter Stability}
The results, obtained by varying hyperparameters, are illustrated in \figref{fig:candidate} and \figref{fig:temperature}. Specifically, \figref{fig:candidate} demonstrates the \textit{Pass@k} across different $n$ values, while \figref{fig:temperature} displays the \textit{Pass@k} for varying inference temperatures.
In both result sets, it is evident that \tool consistently and significantly diminishes the performance of the code generation model across a diverse range of $n$ settings and temperature configurations.




\begin{center}
\begin{tcolorbox}[colback=blue!8,
                  colframe=black,
                  width=\textwidth,
                  arc=1mm, auto outer arc,
                  boxrule=0.9pt,
                 ]
\small Answers to \textbf{RQ4}: \tool consistently delivers stable benchmarking results across random trials and a wide array of hyperparameters. \normalsize

\end{tcolorbox}
\end{center}

%% file: Table/stability.tex
\begin{table}[htbp]
  \centering
  \caption{The stability results of multiple random trials}
  \vspace{-3mm}
  \resizebox{0.88\textwidth}{!}{
    \begin{NiceTabular}{l|l|c|cccccccccccc}
    \CodeBefore
        \rowcolors{2}{}{blue!8}
        \Body
    \toprule
    \toprule
    \multirow{2}[4]{*}{\textbf{Dataset}} & \multirow{2}[4]{*}{\textbf{Methods}} & \multirow{2}[4]{*}{\textbf{Trial ID}} & \multicolumn{3}{c}{\textbf{CodeGen-2b}} & \multicolumn{3}{c}{\textbf{CodeGen2-3.7b}} & \multicolumn{3}{c}{\textbf{Incoder-6b}} & \multicolumn{3}{c}{\textbf{Santacoder-1.1b}} \\
\cmidrule{4-15}          &       &       & \textbf{Pass@1} & \textbf{Pass@10} & \textbf{Pass@100} & \textbf{Pass@1} & \textbf{Pass@10} & \textbf{Pass@100} & \textbf{Pass@1} & \textbf{Pass@10} & \textbf{Pass@100} & \textbf{Pass@1} & \textbf{Pass@10} & \textbf{Pass@100} \\
    \midrule
    \multirow{12}[4]{*}{\textbf{Human-Eval}} & \multirow{6}[2]{*}{\textbf{PPM-V}} & \textbf{1 } & 0.03  & 0.10  & 0.21  & 0.01  & 0.06  & 0.12  & 0.02  & 0.08  & 0.17  & 0.02  & 0.06  & 0.12  \\
          &       & \textbf{2 } & 0.03  & 0.10  & 0.19  & 0.01  & 0.06  & 0.14  & 0.01  & 0.05  & 0.11  & 0.02  & 0.07  & 0.14  \\
          &       & \textbf{3 } & 0.02  & 0.10  & 0.18  & 0.01  & 0.05  & 0.11  & 0.02  & 0.07  & 0.13  & 0.02  & 0.07  & 0.14  \\
          &       & \textbf{4 } & 0.02  & 0.09  & 0.17  & 0.01  & 0.04  & 0.11  & 0.01  & 0.06  & 0.17  & 0.02  & 0.07  & 0.16  \\
          &       & \textbf{5 } & 0.02  & 0.10  & 0.25  & 0.00  & 0.03  & 0.09  & 0.02  & 0.08  & 0.16  & 0.02  & 0.07  & 0.16  \\
          &       & \textbf{Avg ± Std} & 0.02 ± 0.003 & 0.10 ± 0.005 & 0.20 ± 0.032 & 0.01 ± 0.005 & 0.05 ± 0.011 & 0.11 ± 0.020 & 0.02 ± 0.005 & 0.07 ± 0.016 & 0.15 ± 0.027 & 0.02 ± 0.003 & 0.07 ± 0.006 & 0.15 ± 0.016 \\
\cmidrule{2-15}          & \multirow{6}[2]{*}{\textbf{PPM-T}} & \textbf{1 } & 0.01  & 0.07  & 0.16  & 0.01  & 0.03  & 0.08  & 0.01  & 0.04  & 0.11  & 0.01  & 0.04  & 0.12  \\
          &       & \textbf{2 } & 0.02  & 0.06  & 0.16  & 0.00  & 0.03  & 0.07  & 0.01  & 0.04  & 0.12  & 0.01  & 0.04  & 0.11  \\
          &       & \textbf{3 } & 0.02  & 0.07  & 0.20  & 0.01  & 0.04  & 0.10  & 0.01  & 0.03  & 0.09  & 0.02  & 0.06  & 0.16  \\
          &       & \textbf{4 } & 0.01  & 0.06  & 0.19  & 0.01  & 0.04  & 0.09  & 0.01  & 0.03  & 0.07  & 0.01  & 0.05  & 0.12  \\
          &       & \textbf{5 } & 0.02  & 0.07  & 0.18  & 0.01  & 0.03  & 0.06  & 0.01  & 0.04  & 0.12  & 0.01  & 0.05  & 0.14  \\
          &       & \textbf{Avg ± Std} & 0.01 ± 0.003 & 0.07 ± 0.004 & 0.18 ± 0.018 & 0.01 ± 0.002 & 0.03 ± 0.007 & 0.08 ± 0.014 & 0.01 ± 0.002 & 0.04 ± 0.003 & 0.10 ± 0.020 & 0.01 ± 0.004 & 0.05 ± 0.009 & 0.13 ± 0.018 \\
    \midrule
    \midrule
    \multirow{12}[4]{*}{\textbf{MBPP}} & \multirow{6}[2]{*}{\textbf{PPM-V}} & \textbf{1 } & 0.03  & 0.17  & 0.36  & 0.01  & 0.07  & 0.19  & 0.02  & 0.10  & 0.28  & 0.03  & 0.15  & 0.34  \\
          &       & \textbf{2 } & 0.03  & 0.17  & 0.37  & 0.01  & 0.06  & 0.17  & 0.02  & 0.10  & 0.27  & 0.04  & 0.14  & 0.30  \\
          &       & \textbf{3 } & 0.04  & 0.18  & 0.37  & 0.01  & 0.07  & 0.20  & 0.02  & 0.10  & 0.28  & 0.03  & 0.15  & 0.34  \\
          &       & \textbf{4 } & 0.03  & 0.16  & 0.39  & 0.01  & 0.07  & 0.21  & 0.02  & 0.10  & 0.27  & 0.03  & 0.14  & 0.36  \\
          &       & \textbf{5 } & 0.04  & 0.18  & 0.37  & 0.01  & 0.07  & 0.18  & 0.02  & 0.11  & 0.27  & 0.03  & 0.14  & 0.32  \\
          &       & \textbf{Avg ± Std} & 0.04 ± 0.003 & 0.17 ± 0.008 & 0.37 ± 0.011 & 0.01 ± 0.002 & 0.07 ± 0.007 & 0.19 ± 0.015 & 0.02 ± 0.001 & 0.10 ± 0.004 & 0.27 ± 0.006 & 0.03 ± 0.002 & 0.14 ± 0.006 & 0.33 ± 0.020 \\
\cmidrule{2-15}          & \multirow{6}[2]{*}{\textbf{PPM-T}} & \textbf{1 } & 0.06  & 0.22  & 0.39  & 0.03  & 0.12  & 0.24  & 0.02  & 0.10  & 0.25  & 0.04  & 0.14  & 0.31  \\
          &       & \textbf{2 } & 0.06  & 0.21  & 0.38  & 0.03  & 0.12  & 0.22  & 0.02  & 0.10  & 0.26  & 0.04  & 0.14  & 0.30  \\
          &       & \textbf{3 } & 0.07  & 0.23  & 0.41  & 0.03  & 0.11  & 0.21  & 0.02  & 0.09  & 0.24  & 0.04  & 0.15  & 0.29  \\
          &       & \textbf{4 } & 0.06  & 0.21  & 0.40  & 0.03  & 0.11  & 0.24  & 0.03  & 0.11  & 0.25  & 0.04  & 0.16  & 0.30  \\
          &       & \textbf{5 } & 0.06  & 0.22  & 0.40  & 0.03  & 0.11  & 0.25  & 0.02  & 0.10  & 0.27  & 0.03  & 0.14  & 0.29  \\
          &       & \textbf{Avg ± Std} & 0.06 ± 0.004 & 0.22 ± 0.007 & 0.39 ± 0.011 & 0.03 ± 0.001 & 0.11 ± 0.003 & 0.23 ± 0.016 & 0.02 ± 0.002 & 0.10 ± 0.007 & 0.25 ± 0.011 & 0.04 ± 0.003 & 0.14 ± 0.007 & 0.30 ± 0.008 \\
    \bottomrule
    \bottomrule
    \end{NiceTabular}%
    }
  \label{tab:stability}%
\end{table}%

%% file: Text/application.tex


%% file: Text/discussion.tex
\section{Threats to Validity}

\fakeparagraph{Internal Threat} 
Our primary internal concern revolves around the absence of a ground truth for assessing problem naturalness. To tackle this, we adopt a comprehensive approach, utilizing three evaluation metrics: \textit{perplexity} to assess real-world relevance, \textit{IDE warnings} to identify typos and grammar issues, and human evaluations.
Additionally, we verify the real-world relevance of our generated problems by evaluating \lcgms' accuracy in solving the newly introduced lambda programming problems. These results confirm the natural and realistic nature of our lambda problems. Then if the seed problem is a natural one, our newly generated problem would be natural.
Another concern pertains to how randomness affects long-term data integrity. we alleviate this threat by the following efforts. First, as \tool consistently provides stable results across multiple trials, thus, developers may choose not to publish the specific dataset on the Internet, but rather focus on publishing the method. With the public availability of the method, we simulate the probability of problem repetition, observing minimal recurrence. Moreover, after 100 \tool attempts, approximately 70\% of the problems remain distinct. Additionally, \tool offers configurability through a private "offset" parameter, making it challenging for others to replicate problems by keeping this space publicly inaccessible.

\fakeparagraph{External Threat} 
Our external concern pertains to the selection of experimental subjects, such as datasets, models, and baselines.
We aim to mitigate this concern through the following measures: (1) The chosen datasets and models are highly popular and extensively utilized in related research. (2) These datasets and models encompass diverse model types, model training algorithms, and data complexities, offering a broad spectrum of variation. (3) The selected baseline methods include almost all existing type of perturbations.
Hence, while specific data might vary for other subjects, our experimental conclusions should generally remain valid due to this comprehensive selection.

%% file: Text/related.tex
\section{Related work}

\fakeparagraph{Code Models for Software Engineering}
In \secref{sec:background}, we discussed code generation models, and now we introduce other code models for software engineering applications.
Code representation models like \textsf{code2vec} \cite{alon2019code2vec} and \textsf{code2seq} \cite{alon2018code2seq} leverage syntax and structure, performing well in downstream tasks \cite{chen2022learning, austin2021program}. Recently, pre-trained NLP  models like \textsf{BERT} \cite{devlin2018bert} and \textsf{GPT-3} \cite{brown2020language} demonstrated strong transferability to Programming Languages (PL), outperforming \textsf{code2vec} and \textsf{code2seq} \cite{ding2022can}. This success has popularized pre-trained code models, benefiting diverse tasks \cite{buratti2020exploring,feng2020codebert,guo2021dawn,kanade2020learning,svyatkovskiy2020intellicode,wang2021codet5}. \textsf{CuBERT} \cite{kanade2020learning} and \textsf{C-BERT} \cite{buratti2020exploring} use \textsf{BERT} architecture, specifically trained on Python and C source code, respectively. However, their single-language training limits applicability in diverse scenarios.



\fakeparagraph{Robustness Evaluation for Code Model}
Recently, several attack algorithms \cite{nie2019adversarial,zhou2022adversarial,wang2021adversarial,nie2019adversarial,henkel2022semantic,zhou2022adversarial,chen2022nmtsloth,chen2022nicgslowdown} have been proposed to assess the robustness of code models. However, most of these existing methods \cite{li2022poison,yang2023stealthy,yang2022natural,campbell2019theory,branch2022evaluating,zhuo2023robustness} primarily target classification code models. For instance, Yefet et al.  presented \textsf{DAMP} \cite{yefet2020adversarial}, a white-box attack technique that adversarially alters variables in code using gradient information from the victim model. 
However, the evaluation of robustness in pre-trained code generation models has received limited research attention. 

%% file: Text/conclusion.tex
\section{Conclusion}

This paper introduces \tool, a novel approach for benchmarking \lcgms by merging programming problems to create diverse datasets automatically. \tool enhances dataset diversity and ensures long-term data integrity by accepting random input values for each trial. Additionally, it employs lambda programming tasks to maintain coherent and linguistically accurate descriptions. Extensive experiments demonstrate \tool's superiority in challenging code generation models, showcasing exceptional diversity and naturalness in the generated problems. 